\documentclass[11pt,aps,prl,floatfix,superscriptaddress,amsmath,amssymb,noshowpacs]{revtex4}
\usepackage{graphics,lscape,graphicx,epsfig,amsmath,amssymb,epstopdf,wasysym}
\allowdisplaybreaks

\newcommand{\be}{\begin{equation}}
\newcommand{\ee}{\end{equation}}
\newcommand{\bea}{\begin{eqnarray}}
\newcommand{\eea}{\end{eqnarray}}
\newcommand{\ba}{\begin{eqnarray*}}
\newcommand{\ea}{\end{eqnarray*}}

\newcommand{\eqn}[1]{(\ref{#1})}

\newcommand{\ep}{{\epsilon}}

\newcommand{\bw}{\begin{widetext}}
\newcommand{\ew}{\end{widetext}}
\newcommand{\meV}{\,\text{meV}}
\begin{document}

\title{Cooling quasiparticles in A$_3$C$_{60}$ fullerides by excitonic mid-infrared absorption}

\author{Andrea Nava} 
\affiliation{International School for
  Advanced Studies (SISSA), Via Bonomea
  265, I-34136 Trieste, Italy} 
\author{Claudio Giannetti} 
\affiliation{Interdisciplinary Laboratories for Advanced Materials Physics (ILAMP),
Universit\`a Cattolica del Sacro Cuore, Brescia I-25121, Italy} 
\author{Antoine Georges}
\affiliation{Centre de Physique Th\'eorique, E\'cole Polytechnique,
CNRS, Universit\'e Paris-Saclay, 91128 Palaiseau, France}
\affiliation{Coll\'ege de France, 11 place Marcelin Berthelot, 75005 Paris, France}
\affiliation{Department of Quantum Matter Physics, University of Geneva, 24 Quai Ernest-Ansermet, 1211 Geneva 4, Switzerland}
\author{Erio Tosatti}
\affiliation{International School for
  Advanced Studies (SISSA), Via Bonomea
  265, I-34136 Trieste, Italy} 
  \affiliation{CNR-IOM Democritos,  Via Bonomea
  265, I-34136 Trieste, Italy} 
\affiliation{International Centre for Theoretical Physics (ICTP), Strada Costiera 11, I-34151 Trieste, Italy}
\author{Michele Fabrizio} 
\affiliation{International School for
  Advanced Studies (SISSA), Via Bonomea
  265, I-34136 Trieste, Italy}   


\maketitle

\textbf{
Long after its discovery 
superconductivity in alkali fullerides A$_3$C$_{60}$ still challenges conventional wisdom. 
The freshest inroad in such ever-surprising physics is 
the behaviour under intense infrared (IR) excitation. Signatures attributable to a transient superconducting state extending up to temperatures ten times higher than the equilibrium $T_c\sim$~20~K 
have been discovered in K$_3$C$_{60}$ 
after ultra-short pulsed IR irradiation -- an effect which still appears as remarkable as mysterious. Motivated by the observation that the phenomenon is observed in a broad pumping frequency range that coincides with the mid-infrared electronic absorption peak still of unclear origin, rather than to TO phonons as has been proposed,  we advance here a radically new mechanism.  
First, we argue that this broad absorption peak 
represents a "super-exciton" involving the promotion of one electron from the $t_{1u}$ half-filled state to a higher-energy empty $t_{1g}$ state, dramatically
lowered in energy by the large dipole-dipole interaction acting in conjunction with Jahn Teller 
effect within the enormously degenerate manifold of $\big(t_{1u}\big)^2\big(t_{1g}\big)^1$ states. Both long-lived and entropy-rich because they are triplets, the IR-induced excitons act as a sort of cooling mechanism that permits transient superconductive signals to persist up to much larger temperatures.
}

\bigskip

Superconducting
alkali doped fullerenes A$_3$C$_{60}$ are molecular compounds where several actors play together 
to determine an intriguing physical behaviour. The high icosahedral symmetry of C$_{60}$ implies, prior to intermolecular hybridisation, a large degeneracy of the molecular orbitals, thus a strong electronic response to JT molecular distortions lowering that symmetry. 
In particular, the $t_{1u}$ LUMO, which accommodates the three electrons donated by the alkali metals, is threefold degenerate and JT 
coupled to eight fivefold-degenerate molecular vibrations of $H_g$ symmetry, which mediate the 
pairing\cite{Gunnarsson-review}. The JT effect, favouring low spin, is partly hindered by  
(Coulomb) Hund's rule exchange, which favours high spin. 
Therefore the overall singlet pairing strength $g$, though still sizeable, is 
way too small compared to the charging energy of each C$^{3-}_{60}$ to justify by simple arguments why A$_3$C$_{60}$ are $s$-wave superconductors. 
The explanation of this puzzle proposed in \cite{Science2002,nostroRMP} and vindicated by recent experiments emphasises the crucial role of a parent Mott insulating state where 
the JT coupling effectively inverts Hund's rules, the molecular ground state therefore turning to
spin $S=1/2$ rather than $S=3/2$~\cite{FabrizioPRB1997}.  A $S=1/2$ antiferromagnetic  insulating phase is indeed the ground state in over-expanded NH$_3$K$_3$C$_{60}$\cite{Durand2003,Kitano} and in
Cs$_3$C$_{60}$\cite{Prassides2012} at ambient pressure. 
In the metallic state, attained under pressure in  Cs$_3$C$_{60}$ and at ambient pressure in 
K$_3$C$_{60}$ and Rb$_3$C$_{60}$,  
the incipient Mott localisation slows down the coherent motion of quasiparticles while undressing them from charge correlations. As a result, the singlet pairing strength $g$ eventually overwhelms the quasiparticle Coulomb pseudopotential and, on approaching the Mott transition, the system is effectively 
driven towards the top of the universal T$_c$ \textit{vs.} $g$ curve 
\cite{Robaszkiewicz}, where the critical temperature reaches the maximum possible value at a given non-retarded attraction
 $T_c^\text{MAX}\sim 0.055\,g$. 
Thus, according to the theory of Ref.~\cite{nostroRMP}, the peak $T_c \sim 38~\text{K}$ reached by Cs$_3$C$_{60}$ at $\sim 7~\text{kbar}$ \cite{Ganin,Alloul2013}
is actually the highest attainable at equilibrium in fullerides.\\
\begin{figure}[thb]
\centerline{\includegraphics[keepaspectratio,clip,width=.8\textwidth,angle=270]{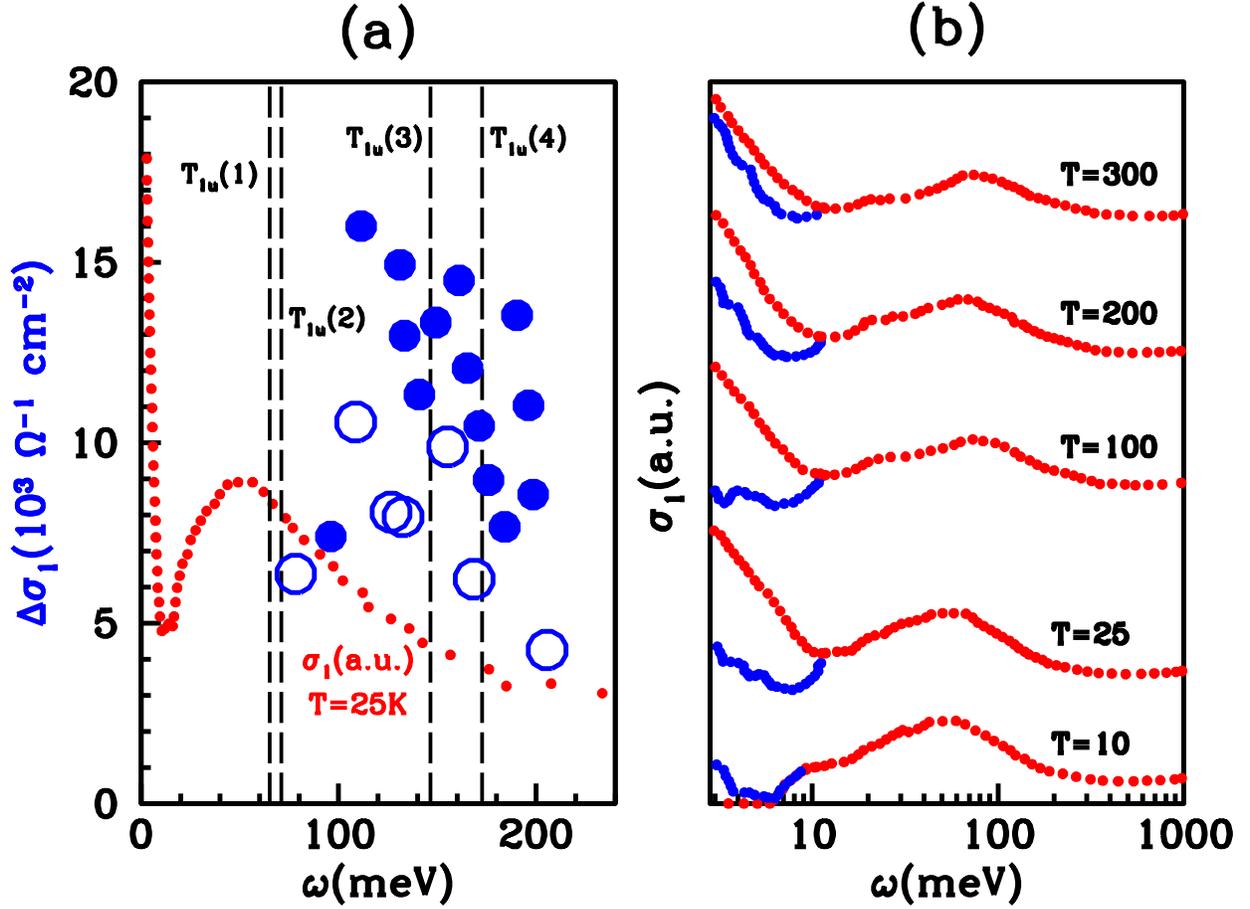}}
\vspace{-0.7cm}
\caption{ \footnotesize{\textbf{Experimental data from Ref.~\cite{Cavalleri2016}. (a) Blue circles: spectral-weight reduction, $\Delta\sigma_1(\omega)$,  in the optical conductivity as a 
function of the laser frequency $\omega$ from different measurements (data from Fig.~4d of Ref.~\cite{Cavalleri2016}). Red dots: equilibrium optical conductivity at 
$T=25~\text{K}$ with the broad IR peak which we interpret as involving a  $(t_{1u})^3 \cup (t_{1u})^2(t_{1g})^1$ exciton.  
Dashed vertical lines: frequencies of the $T_{1u}$ TO phonon modes. Note that, except at two lowest frequency points, 
$\Delta\sigma_1(\omega)$  follows closely the shape of the mid IR absorption peak, rather than peaking at the $T_{1u}$ frequencies. 
(b) Optical conductivities at equilibrium, red dots, and 1~ps after the photoexcitation, blue dots, for different temperatures, below, $T=10~\text{K}$, and above, 
$T=25,100,200,300~\text{K}$ the equilibrium $T_c$  (data from Figs. 2b, 2e, 3b, 3e and from Extended data figure 9e of Ref.~\cite{Cavalleri2016})} }}
\label{Fig-1}
\end{figure}
This equilibrium upper limit has been far surpassed in out-of-equilibrium conditions in a recent
remarkable pump-probe experiment on K$_3$C$_{60}$\cite{Cavalleri2016}. 
After irradiation by an intense femtosecond infrared pulse between 80 and 200 meV, K$_3$C$_{60}$ showed a transient regime of some picoseconds where the optical properties looked like those of a superconductor, alas up to a temperature $T\gtrsim 200$~K, 
ten times higher than the equilibrium $T_c\sim 20$~K, see Fig.~\ref{Fig-1}(b). This tantalising 
observation has already elicited various theoretical efforts \cite{GeorgesPRB2016,Demler2016,Georges2016-2,Millis2016,Mazza2017}, 
where it was mainly assumed, as in the original work \cite{Cavalleri2016}, 
that TO phonon IR absorption acts as the crucial ingredient   
increasing the pairing efficiency. 
Here we follow another route directly inspired by experimental features, which leads to a totally different perspective.\\

First of all, the transient "superconducting" gap does increase~\cite{Cavalleri2016}, yet 
not as much as the transient $T_c$, see Fig.~\ref{Fig-1}(b). 
More importantly, we note in Ref.~\cite{Cavalleri2016} that the transient reduction of optical conductivity (suggestive of a transiently enhanced superconducting state) is broadly distributed over the IR 
pumping frequency range from 80 to 200 meV, see Fig.~\ref{Fig-1}(a). 
Although that includes the 
two highest $T_{1u}$ IR-active modes near 150 and 170 meV\cite{Mihaly1993}, the enhancement does not especially peak there,  extending
instead to lower frequencies, see Fig.~\ref{Fig-1}(a). There is instead an intriguing similarity between a long known~\cite{DegiorgiPRB1994,Degiorgi-ADP1998} broad absorption peak that characterises the 
equilibrium IR response of K$_3$C$_{60}$ and Rb$_3$C$_{60}$. 
This peak is present and strong in the equilibrium optical data of Ref.~\cite{Cavalleri2016}, 
centred around $\sim$ 50 meV and $\sim$ 100 meV broad, see Fig.~\ref{Fig-1}(b). Given 
these characteristics, the underlying excitation is not a phonon, and can only be electronic; yet, nobody seems to know exactly 
what it is~\cite{Gelfand1993,DeshpandePRB1994,Gunnarsson&EyertPRB1998,Chibotaru-Meroedrico}.

Intriguingly, it now appears that the superconducting enhancement follows rather closely the shape of this 
IR absorption feature. 
Our first task is therefore to understand this excitation which
might provide a precious clue to superconductivity enhancement in alternative to the 
resonance with infrared-active TO modes.\\
In A$_3$C$_{60}$ the conduction electrons occupy the narrow band originated by the threefold degenerate $t_{1u}$ LUMO of C$_{60}$. The Coulomb interaction projected onto the $t_{1u}$ manifold includes a charge repulsion, the Hubbard $U \sim 1~\text{eV}$,  plus a quadrupole-quadrupole electronic interaction providing an intra-molecular Hund's rule exchange $J_\text{H}>0$. The latter splits the twenty possible $(t_{1u})^{3}$ configurations of C$_{60}^{3-}$, assumed at first   
with nuclei rigidly frozen in their ideal icosahedral positions, as 
\be
\begin{split}
E\left(^2T_{1u}\right) - E\left(^4\!A_{u}\right) &= 10\,J_\text{H}\,,\\
E\left(^2H_{u}\right) - E\left(^4\!A_{u}\right) &= 6\,J_\text{H}\,.
\end{split}
\ee
The highest-spin state, $^4\!A_{u}$, has therefore the lowest energy, see Table~\ref{Table1}. 
Once the nuclei 
defreeze, and the molecular ion can distort, the resulting JT energy $E_{JT}$ strongly competes against exchange $J_\text{H}$ ,
since now the quadrupole operators of the $t_{1u}$ electrons couple with the quadrupole of the $H_g$ vibrational modes, but with opposite sign.~\cite{nostroRMP}
In C$_{60}^{3-}$, the JT effect actually prevails over Coulomb exchange, effectively inverting Hund's rules. The real ground state 
thus becomes 
the low-spin $^2T_{1u}$  multiplet\cite{ManiniPRB94-1,ManiniPRB94-2, O'BrienPRB96,DunnPRB2005,SigristPRB2007,Shahab2016}.\\
\begin{table}
\begin{tabular}{|c|c|c|}\hline
~~E(meV)~~ & ~~$(t_{1u})^3$~~ & ~~$(t_{1u})^2(t_{1g})^1$~~ \\ \hline\hline
0   & $^4A_u$       &    \\ \hline
285    & $^2H_u$       &\\ \hline
476    &  $^2T_{1u}$             & \\ \hline 
494  &     &$^4H_g$\\ \hline
525    & 				& $^2T_{1g}$ \\ \hline
618   &				& $^4A_g$ \\ \hline
1109    &				& $^2G_g+^2T_{2g}$\\ \hline
1143    &				& $^4T_{1g}$ \\ \hline
1280   &				& $^2H_g$ \\ \hline
1496   &				& $^2T_{1g}$ \\ \hline
1947   &				& $^2H_g$ \\ \hline
2218   &				& $^2A_g$\\ \hline
2549   & 				& $^2T_{1g}$ \\ \hline
\end{tabular}
\caption{\footnotesize{\textbf{Molecular terms of {\it undistorted} C$_{60}^{3-}$ 
calculated 
in the absence of Jahn Teller coupling within the $(t_{1u})^3 \cup 
(t_{1u})^2(t_{1g})^1$ manifold according to the interaction model III of Ref.~\cite{Nikolaev&Michel}. The zero of energy is set at the lowest $^4\!A_u$ state and the  
single-particle energy difference between $t_{1g}$ and $t_{1u}$, 
$\Delta\ep = 1153\meV$, is the same as in \cite{Nikolaev&Michel}. The electronic configurations are labeled according to the irreducible representations of the icosahedral group $I_h$. Since the $t_{1u}$ and $t_{1g}$ orbitals transform as atomic $p$ orbitals, they are in one to one correspondence with angular momentum states in $O(3)$ symmetry: $A\to S$, $T\to P$, $H\to  D$ and 
$G+T \to F$. Note the exceptionally wide spread of terms in the $(t_{1u})^2(t_{1g})^1$ configuration.
 }}}
\label{Table1}
\end{table}
Next, what about the  $(t_{1u})^2(t_{1g})^1$ configuration? 
Within each C$_{60}^{3-}$ molecule, the lowest dipole-allowed excitation corresponds to transferring one electron from the $t_{1u}$ LUMO to the $t_{1g}$ LUMO+1, which is also threefold degenerate and whose single-particle energy level lies $\Delta\ep \sim 1.2~\text{eV}$ above. High as this energy is, 
the $(t_{1u})^2\,(t_{1g})^1$ subspace comprises as many as
90 states, hence many times more susceptible to exchange splitting and JT effects 
than the lowest energy $(t_{1u})^{3}$ subspace. In addition,  the Coulomb interaction projected onto the enlarged $t_{1u}$--$t_{1g}$ manifold also includes a dipole-dipole interaction,
which is  stronger than the quadrupole-quadrupole. 
Through a fully quantitative multipole expansion of the Coulomb interaction,  Nikolaev and Michel found  (omitting JT couplings)\cite{Nikolaev&Michel}  
that the split $(t_{1u})^2\,(t_{1g})^1$ subspace spans a gigantic $2\,\text{eV}$ range,
four times wider than the  
splitting $10J_\text{H}\simeq 476\meV$ of the $(t_{1u})^{3}$, see Table~\ref{Table1}. 
\begin{figure}[thb]
\centerline{\includegraphics[keepaspectratio,clip,width=.8\textwidth,angle=270]{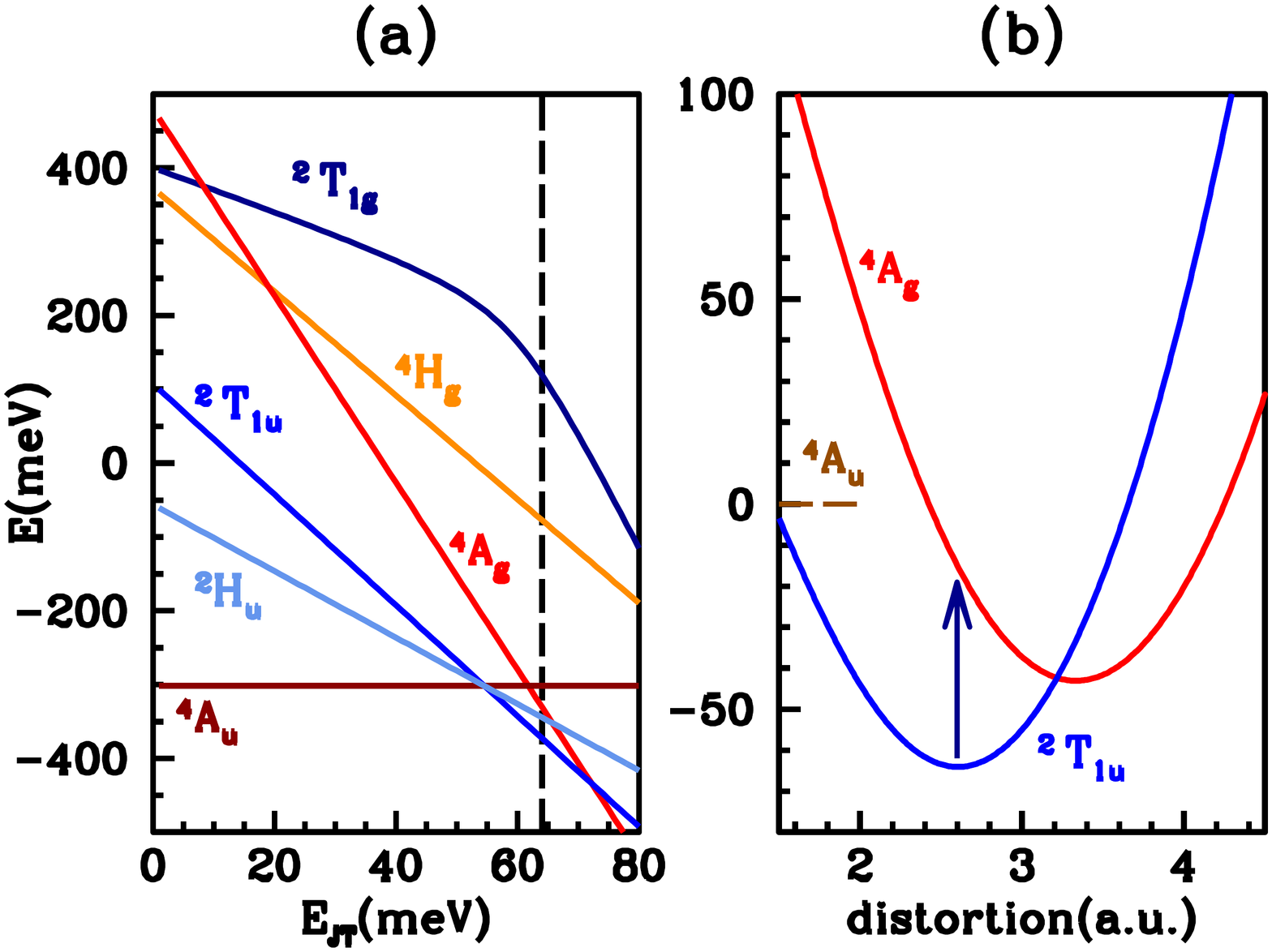}
}
\vspace{-0.5cm}
\caption{\footnotesize{\textbf{Molecular terms in the presence of Jahn-Teller. Panel (a): Low-lying C$_{60}^{3-}$ molecular terms as function of the $t_{1u}$ JT energy $E_\text{JT}$ in the antiadiabatic approximation. 
Ungerade ($u$) and gerade ($g$) terms derive from   $(t_{1u})^3 $  and from $(t_{1u})^2(t_{1g})^1$
configurations, respectively. 
Terms are calculated as in Table~\ref{Table1}, now with effective exchange parameters 
including JT contributions evaluated in the anti-adiabatic approximation (see Supplementary Notes). 
The Coulomb exchange parameters such as $J_\text{H}$ are the same as in Table~\ref{Table1} with a 14\% reduction to mimic screening effects. The $t_{1g}$ JT energy is taken as $1.25^2\,E_\text{JT}$ and $\Delta\ep=1240\meV$ to account for the overestimate of the $t_{1g}$ downward single-particle energy-shift within the antiadiabatic approximation.
The vertical dashed line indicates the suggested  appropriate parameter for  K$_3$C$_{60}$.
Panel (b): Energy as a function of the modulus of the JT distortion of the $^2T_{1u}$ and $^4\!A_g$ configurations. The zero of energy is set at the $^4\!A_u$ level
and we take $\Delta\ep=1080\meV$ and an interaction screening reduction of 22\%. The calculation is performed within the single mode approximation by the variational approach of Ref.~\cite{SigristPRB2007} using a 
mode frequency $\omega=100\meV$, vibrational coupling $g=1.32$ for $t_{1u}$, which corresponds to $E_\text{JT}=87\meV$, and $1.25\,g$ for $t_{1g}$. The differences in screening reduction and $\Delta\ep$ with respect to the left panel take into account the overestimated JT effect within the antiadiabatic approximation. The arrow shows
the vertical Franck-Condon exciton transition.
}}}
\label{Fig-2}
\end{figure}
The two lowest $(t_{1u})^2\,(t_{1g})^1$
states with symmetry $^4H_g$ and $^2T_{1g}$ lie at only $494\meV$ and $525\meV$, respectively, above the $^4\!A_{u}$ ground state~\cite{Negri1992}, 
and that is before JT coupling. 
After allowing for JT, there is a further lowering, and
the situation becomes richer~\cite{Rai1996,ChibotaruPRB1996}. The quadrupole moment of the $t_{1g}$ LUMO+1 has opposite sign to the $t_{1u}$ LUMO, and its absolute value is 2.6 times larger,  which
makes JT couplings much more effective. 
In particular, and unlike the $(t_{1u})^3$ manifold, 
the JT effect in $(t_{1u})^2\,(t_{1g})^1$ 
is stronger in the high-spin $S=3/2$ subspace than in low-spin $S=1/2$. The reason is that  
in the $S=3/2$ subspace the $H_g$ vibrations couple 
together the lowest energy $^4H_g$  
with the $^4A_g$ term, which is a mere $124\meV$ above, see Table~\ref{Table1}. 
In the $S=1/2$ subspace, by contrast, the lowest energy $^2T_{1g}$ is only coupled to states higher than $600\meV$ above, 
which reduces the effect. \\ 
We further note that the new $t\otimes H$  $(t_{1u})^2\,(t_{1g})^1$ JT problem within configurations $^4H_g$ and $^4A_g$ is equivalent to that of $C_{60}^{2-}$ in the $S=0$ subspace of the $(t_{1u})^2$ manifold, which involves 
the configurations $^1H_g$ and $^1A_g$ and where the JT energy gain is known to be maximum
\cite{ManiniPRB94-1,ManiniPRB94-2,O'BrienPRB96,SigristPRB2007}.   
On the other hand, the Coulomb exchange splitting $E(^4A_g)-E(^4H_g)=124\meV$  
of $(t_{1u})^2\,(t_{1g})^1$ $S=3/2$ subspace is smaller than $E(^1A_g)-E(^1H_g)=6J_\text{H}=286\meV$ of the $S=0$ $(t_{1u})^2$ case, 
implying a larger JT energy gain.\\  
\begin{figure}[thb]
\centerline{\includegraphics[keepaspectratio,clip,width=.9\textwidth]{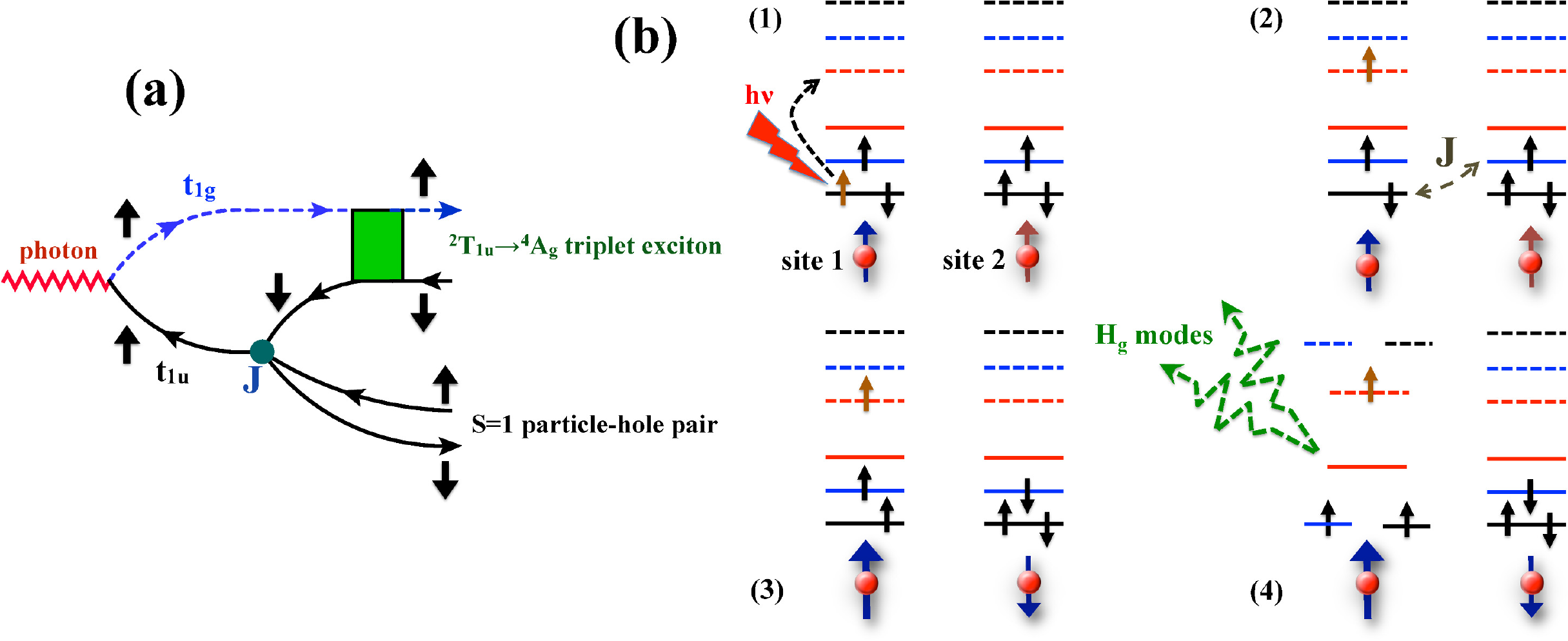}}
\caption{\footnotesize{\textbf{The absorption process. Panel (a): diagrammatic representation of the absorption. The photon induces a virtual dipolar transition 
$t_{1u}\to t_{1g}$. This intermediate state 
spawns a triplet super-exciton and a spin-triplet particle-hole pair.  The high density of the latter near a Mott transition
boosts the strength of this process.  The circular vertex $J$ is the \textit{intermolecular} four-leg vertex exchange, the square electron-hole interaction
comprises intramolecular exchange, JT and Franck-Condon effects, and all $t_{1u}$ (solid black lines) and $t_{1g}$ (dashed blue ones) Green's functions include self-energy corrections.
Panel (b): pictorial representation of the same process with the absorption of a paramagnon. In the initial state (1) two sites are both in the $^2T_{1u}$ configuration, here drawn in the static JT limit. 
Note that site 2 is in a spin-state disfavoured by the inter-site antiferromagnetic exchange, i.e. a paramagnon excitation is assumed to be present. The photon with energy $h\nu$ transfers a $t_{1u}$ electron (solid levels) into the $t_{1g}$ orbitals (dotted levels). Note that the JT distortion of the $t_{1u}$ orbitals is opposite to that of $t_{1g}$, as highlighted by the colours of the orbitals. In (2) $\to$ (3) the antiferromagnetic exchange $J$ flips the spins of the two sites, 1 and 2, so that site 2 has now the 
right antiferromagnetic spin direction: the paramagnon has been absorbed. Through the emission of $H_g$ vibrations, site 1 relaxes to the optimal JT distortion corresponding to the state $^4\!A_g$. 
}}}
\label{Fig-3}
\end{figure}
Accurate estimates of the molecular terms within the enlarged $(t_{1u})^3\cup(t_{1u})^2(t_{1g})^1$ manifold and in presence of JT coupling to the $H_g$ vibrations would
require a precise knowledge of all Hamiltonian parameters that are involved. That's a tall order, because, while the frequencies of the $H_g$ modes are known from 
experiments, and different calculations of JT energy $E_{JT}$ more or less agree,  the individual values of the coupling constants with 
the $t_{1u}$ electrons are hard to establish~\cite{IwaharaPRB2010} without resorting to photoemission experiment~\cite{Gunnarsson1995}.  Also questionable is
whether the simple linear coupling to vibrations, as usually assumed, is sufficient, as has been pointed out~\cite{BunnPRB2008,DunnPRB2013}. Besides that, 
there are so far no direct estimate of the $H_g$ vibration coupling constants with the $t_{1g}$ electrons, obviously not extractable from photoemission. We mentioned that the $t_{1g}$ quadrupole moment is larger in absolute value than the $t_{1u}$ one, which would suggest stronger vibration coupling constants, as indeed observed in electronic structure calculations of isolated molecular anions~\cite{Green1996}. Moreover, $t_{1g}$ electrons couple preferentially 
to higher frequency $H_g$ vibrations with tangential character, while $t_{1u}$ electrons to lower frequency radial vibrations, which might also imply a larger $t_{1g}$ JT energy~\cite{BunnPRB2008}.   
One should finally note that, given the large size of the 
$(t_{1u})^2(t_{1g})^1$ subspace, even small variations of the many Coulomb exchange parameters\cite{Nikolaev&Michel} and vibrational coupling constants may lead to appreciably different results. \\
For these reasons we opt for a less ambitious approach and, following Ref.~\cite{nostroRMP}, we treat 
the JT problem within the anti-adiabatic approximation, 
were all effects depend only on the value of the total JT energy gain $E_\text{JT}$,
whose value for $t_{1u}$ electrons is far less uncertain than the value of each vibrational coupling constant~\cite{IwaharaPRB2010}, see the Supplementary Notes for details. 
We use the model III interaction parameters of Nikolaev and Michel\cite{Nikolaev&Michel}, with a 14\% reduction to account for screening effects of nearby molecules\cite{GunnarssonPRL92}, and we further assume, in accordance with the density functional  
results of~\cite{Green1996}, that the $t_{1g}$ LUMO+1 JT energy is $1.25^2$ larger than the $t_{1u}$ LUMO one.  
In the left panel of Fig.~\ref{Fig-2} we show the low lying molecular terms as function of $E_\text{JT}$ \cite{IwaharaPRB2010}. \\
We can now consider the full multiplet spectrum for a realistic estimate of $E_\text{JT} = 50 - 70\meV$~\cite{IwaharaPRB2010}.  
The $^2T_{1u}$ ground state and the lowest $^2H_u$ excitation, 
whose role was recently discussed~\cite{Shahab2016}, both belong to the $(t_{1u})^{3}$ manifold. The very next state however is the $^4\!A_g$ term, of $(t_{1u})^2\,(t_{1g})^1$ origin,  dramatically pushed 
down  close to the ground state by JT 
and dipole-dipole interaction, despite the 1~eV energy of the $t_{1g}$ LUMO+1. We also performed a different calculation, treating the JT coupling within the single mode 
approximation~\cite{ManiniPRB98} and using a variational 
approach~\cite{SigristPRB2007} that consists of a statically distorted wavefunction projected onto a state with well defined icosahedral symmetry (details are in the Supplementary Notes). In the right panel of Fig.~\ref{Fig-2} we show the energies thus obtained of the $^2T_{1u}$ and $^4\!A_g$ 
states as function of the distortion vector norm. As anticipated, the $^4\!A_g$ energy minimum is reached for a larger distortion than that of
$^2T_{1u}$, which entails substantial Franck-Condon effects - further strengthened by the shape difference, bimodal for $^2T_{1u}$~\cite{ManiniPRB94-1,ManiniPRB94-2} and unimodal 
for $^4\!A_g$.\\
We propose that the IR peak observed in 
A$_3$C$_{60}$ corresponds precisely to the low lying $^4\!A_g$ state, 
the  $^2T_{1u} \rightarrow ^4\!A_g$ transition essentially turning 
into a genuine triplet exciton in the bulk material. The parity allowed but spin forbidden optical creation of this exciton can actually acquire oscillator strength and
appear in the IR optical spectrum of a narrow-band nearly (antiferro)magnetic metal, through 
the simultaneous absorption/emission of a low energy spin-triplet particle-hole excitation, that is a paramagnon.
For that it is important to recall that A$_3$C$_{60}$ are indeed narrow quasiparticle-band metals, close to a transition into an antiferromagnetic Mott insulator state, so much so that the transition is realised when \textit{the cation A} merely changes from Rb to Cs. The absorption process is schematically shown in Fig.~\ref{Fig-3}. The photon induces a virtual spin-conserving transition $t_{1u}\to t_{1g}$. This intermediate state then transforms into the triplet exciton by absorbing/emitting a paramagnon via intermolecular exchange. One should
note that this absorption mechanism is of the very same
nature to that introduced by Rice and Choi~\cite{Rice&Choi}, which is necessary to explain
why uncharged $T_{1u}$ vibrations acquire oscillator strength and thus are observed in optics. The contribution of the $^2T_{1u} \rightarrow (^4\!A_g$  $\pm$ paramagnon) peak to the optical conductivity reads 
\be
\begin{split}
\delta\sigma_1(\omega) &\propto  
\int_0\! d\ep\,\mathcal{A}_\text{exc}(\ep)\,\Bigg[\,
\theta(\ep-\omega)\,b(\ep-\omega)\,
\chi"(\ep-\omega)+\theta(\omega-\ep)\,
\Big(1+b(\omega-\ep)\Big)\chi"(\omega-\ep)\\
&\phantom{\propto  
\int_0\! d\ep\,\mathcal{A}_\text{exc}(\ep)\,\Bigg[}
\quad - b(\ep+\omega)\,\chi"(\ep +\omega)\,\Bigg]\,,
\end{split}\label{delta-sigma}
\ee
where $\mathcal{A}_\text{exc}(\ep)$ is the exciton absorption spectrum, $b(\ep)$ the Bose distribution function, 
and $\chi"(\ep)$ the imaginary part of the dynamical local spin susceptibility. Equation \eqn{delta-sigma} suggests that 
the large width of the absorption peak, which experimentally corresponds to a timescale of about $7~\text{fs}$, is the result of a convolution between the paramagnon bandwidth and a Franck-Condon broadening, rather than a radiative lifetime of the exciton.
In fact, the expectedly strong Franck-Condon effect must cause 
a large broadening in $\mathcal{A}_\text{exc}(\ep)$ corresponding to the non-radiative relaxation of the triplet exciton to a dark state whose lifetime might be much longer, possibly picoseconds or more,
before eventual (phosphorescent) recombination.  
In agreement with this exciton-paramagnon interpretation, the IR absorption peak grows in importance and intensity from K$_3$C$_{60}$ to Rb$_3$C$_{60}$\cite{DegiorgiPRB1994},  
the latter closer to Mott insulation (realised in Cs$_3$C$_{60}$), thus with stronger and narrower paramagnons.  
\\
The next and central question in the present context is if and why this exciton peak should actually play a role in the apparent enhancement of T$_c$ found by Ref.~\cite{Cavalleri2016}
where IR-pumping is roughly in the same frequency range. 
We start by noting that the 
experimental transient superconducting-like absorption spectra suggest, see Fig.~\ref{Fig-1}(b), 
that the IR pump can act to sweep away the thermally excited 
quasiparticle states that, at equilibrium, are responsible for the gap filling-up and closing with the transition to the normal state. Things superficially seem as if the pump effectively cooled down quasiparticles. 
Following this hypothesis, we can 
qualitatively describe how the quasiparticle distribution should evolve first during the IR laser pulse, about 300~fs long. 
Within that short time lapse, the system is effectively isolated from the environment, with which it was in thermal equilibrium before the IR shot. 
\begin{figure}[thb]
\centerline{\includegraphics[keepaspectratio,clip,width=.7\textwidth,angle=270]{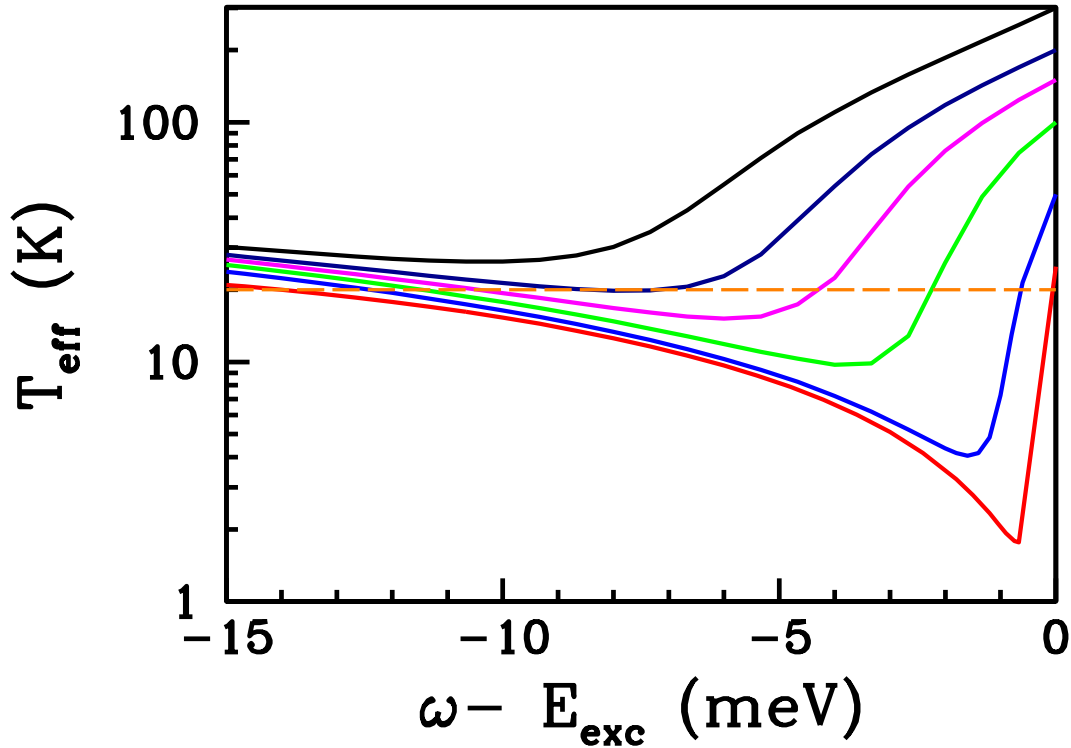}}
\caption{\textbf{Effective temperature $T_\text{eff}$ after the laser pulse calculated as explained in the main body of the text, for initial values $T=300,200,150,100,50,25$~K, from the top curve (black) to the bottom one (red),  
as function of $\omega-E_\text{exc}$, where $\omega$ is the light frequency, with an assumed quasiparticle bandwidth of 100~meV and a sharp exciton line. 
The coefficient that multiplies the process in Fig.~\ref{Fig-3} is fixed to reproduce in linear response the excitonic peak value of the optical conductivity at equilibrium. The horizontal line at 20~K is the value of the equilibrium T$_c$}}
\label{Fig-4}
\end{figure}
The IR pulse supplies the initial normal metal with energy, which is sunk in the exciton-paramagnon 
excitation as well as by the vibrations 
that are emitted during the molecular relaxation after the vertical Franck-Condon transition. 
If we assume that the quasiparticle collision rate is high enough, as expected by the poor Fermi-liquid character above 
T$_c$ \cite{nostroRMP}, then the quasiparticle subsystem will exit the laser shot time in an effective microcanonical 
ensemble identified by an energy $\mathcal{E}$ and quasiparticle number $\mathcal{N}$.
At a later time the quasiparticles will eventually come to equilibrium 
with the excitons, the lattice and the molecular vibrations (the decay times of the eight 
$H_g$ modes into $t_{1u}$ particle-hole excitations range between 0.03 and 4~ps). Yet, in the long transient before that happens, we can legitimately define an entropy 
$\mathcal{S}(\mathcal{E},\mathcal{N})$ of the 
quasiparticle liquid and its effective temperature $T_\text{eff}^{-1} = \partial\mathcal{S}/\partial \mathcal{E}$. Moreover, if the quasiparticle collision integral is strong enough to establish local equilibrium during the whole pulse duration, we are additionally allowed to define an 
entropy $S\big(\mathcal{E}(t),\mathcal{N}(t)\big)$ that depends on the quasiparticle 
energy, $\mathcal{E}(t)$, and number, $\mathcal{N}(t)$, at time $t$ after the pulse front arrives. The absorption process of Fig.~\ref{Fig-3} implies that the creation rate of excitons is  
\be
\dot{N}_\text{exc}(t) = \int d\ep\,\mathcal{A}_\text{exc}(\ep)\,
\dot{n}_\text{exc}(\ep,t)\,,\label{N_exc}
\ee
where $\dot{n}_\text{exc}(\ep,t)$ is equal to the term in square brackets of Eq.~\eqn{delta-sigma} multiplied by a parameter that we fit from equilibrium optical data (see Supplementary Notes), with the Bose distribution function and magnetic susceptibility corresponding to the instantaneous local equilibrium conditions. Since at $\omega \leq E_\text{exc}$, for each extra exciton a quasiparticle is annihilated then $\dot{N}_\text{exc}(t)=-\dot{\mathcal{N}}(t)$. Moreover energy conservation implies that 
\be
\dot{\mathcal{E}}(t) = \int d\ep\,\big(\omega-\ep\big)\,\mathcal{A}_\text{exc}(\ep)\,\dot{n}_\text{exc}(\ep,t)
= \big(\omega-E_\text{exc}\big)\,\dot{N}_\text{exc}(t)
\,,\label{dot-E}
\ee
where $\omega$ is the laser frequency and the last equivalence holds if $\mathcal{A}_\text{exc}(\ep)\sim \delta(\ep-E_\text{exc})$, which we shall assume hereafter for simplicity. 
Because of our assumption of local equilibrium,  it follows that the quasiparticle entropy satisfies 
\be
T(t)\,\dot{\mathcal{S}}(t) = \dot{\mathcal{E}}(t) - \mu(t)\,\dot{\mathcal{N}}(t)
=\int d\ep\,\big(\omega-\ep+\mu(t)\big)\,\mathcal{A}_\text{exc}(\ep)\,\dot{n}_\text{exc}(\ep,t)
= \Big(\omega-E_\text{exc}+\mu(t)\Big)\,\dot{N}_\text{exc}(t)
\,,\label{dot-S}
\ee
where $T(t)$ and $\mu(t)=-T(t)^{-1}\big(\partial \mathcal{S}/\partial \mathcal{N})_\mathcal{E}$
are, respectively, the instantaneous temperature and 
chemical potential. The entropy is expected to be maximum 
when the number of $t_{1u}$ quasiparticles is equal to its initial value of three per molecule, so that $\mu(t)<0$ for any $\mathcal{N}(t)<\mathcal{N}(0)$. Through Eq.~\eqn{dot-S} we thus reach the conclusion that the quasiparticle entropy can indeed decrease, and so the effective temperature, especially for frequencies 
$\omega \leq E_\text{exc}$ when exciton creation requires absorption of 
thermal quasiparticle-quasihole triplet pairs. 
To simplify the calculation of $T_\text{eff}=T(t\simeq 300~\text{fs})$ at the end of the laser pulse, besides assuming $\mathcal{A}_\text{exc}(\ep)\sim \delta(\ep-E_\text{exc})$,  
we also neglect the contribution from 
the change in quasiparticle density, i.e. we take $\mu(t)=0$ in \eqn{dot-S}, which implies that the entropy may decrease only below resonance. Furthermore we assume for $\chi''(\ep)$ the expression of non-interacting quasiparticles at half-filling  
and temperature $T(t)$ with a reduced bandwidth of 100~meV, and model the evolution of their distribution function by a Boltzmann type of equation (see Supplementary Notes). 
In Fig.\ref{Fig-4} we show $T_\text{eff}$ thus obtained for equilibrium sample temperatures 
$T=25,50,100,150,200,300$~K. The result of this modelling, crude but we believe inevitable, 
is that the effective temperature $T_\text{eff}$ can indeed be substantially lower than the equilibrium value 
-- thermal triplet quasiparticle-quasihole pairs being absorbed so that IR pumping can reach the exciton energy.  \\
\textbf{Conclusions} --  The apparently tenfold critical temperature enhancement discovered by IR pumping in K$_3$C$_{60}$ \cite{Cavalleri2016} is explained
by a novel mechanism. 
First,  noting that the effect broadly overlaps in frequency with the unexplained equilibrium mid-infrared absorption peak observed in all A$_3$C$_{60}$ fullerides,
that peak is argued, on the basis of single-molecule calculations, 
to correspond to the creation of a triplet exciton,  Frank-Condon broadened and downshifted  from its high LUMO--LUMO+1 energy by large intra-molecular interactions. Spin conservation requires this process to be accompanied by absorption/emission of a paramagnon.  \\
Second, we propose that the 
transient $T_c$ enhancement occurs because, in the process of promoting quasiparticles into 
these long-lived triplet excitons, the laser pulse effectively cools down the quasiparticles system. 
This also explains why the experiment at 300~K~\cite{Cavalleri2016} still shows a transient increase of reflectivity, even though the optical data cannot be fit by a model for a superconducting state. 
\\ 
Differently from other laser cooling techniques~\cite{KETTERLE1996}, this mechanism relies on the triplet excitons generated by the laser pulse, which effectively act as charge and spin reservoir soaking up entropy from quasiparticles~\cite{GeorgesPRA2009}.\\
While compatible with existing data, various aspects and implications of the present theory can be tested against further experiments. For one, the exciton and its spin-triplet nature could be tackled by magnetic fields and other spectroscopic tools, including e.g., detection of phosphorescence in pumped Cs$_3$C$_{60}$. \\
The possible existence and detection at ambient pressure of the same broad IR absorption peak near 50~meV in the Mott insulating 
A15-Cs$_3$C$_{60}$ at ambient pressure would provide support to our proposal of a light-induced intra-molecular exciton $^4\!A_g$ without charge transfer among nearby molecules Ñ as opposed 
to the alternative $^4\!A_u$, a term which besides spin is also parity forbidden and thus much weaker as it requires additional inter-molecular excitations. It may be noted, on the other 
hand, that the lack of inversion symmetry in merohedrally disordered fcc fullerides might partly allow the parity forbidden dipole transitions~\cite{DeshpandePRB1994,
Chibotaru-Meroedrico}, mixing in case the spin-quartet $^4\!A_u$ state with the $^4\!A_g$. Our theory of pumping-induced cooling is sufficiently general and would apply to that case too.\\ 
Also important would be a re-examination of NMR data, where signatures of a $75\meV$ spin-gap in Rb$_3$C$_{60}$~\cite{AlloulPRB2002} have been so far attributed to thermal population of the $^4A_u$ state, for the possible presence of another, possibly even lower energy $^4A_g$ 
spin-quartet state. \\ 
Finally, the  role of the triplet exciton in the IR-pumping enhancement  of T$_c$ could be addressed in a variety of ways and of materials. The strongest candidate 
remains pressurised Cs$_3$C$_{60}$, which metallizes and superconducts above 5~kbar, and where the full range of parameters becomes available as a function of pressure. 
The ideal maximum equilibrium $T_c$=38 K of fullerides being achieved near 7~kbar\cite{Ganin}, it would be exciting to explore whether the transient $T_c$ might conceivably even be raised closer to room temperature. 
\section*{Acknowledgments}
We are very grateful to A. Cavalleri, L.~F. Chibotaru,   
M. Capone, and A. Cantaluppi for comments and discussions, and to S.S. Naghavi for his help.  We also acknowledge 
discussions with A. Isidori, M. Kim and G. Mazza. This work was supported by the European Union,  
under ERC FIRSTORM, contract N. 692670, ERC MODPHYSFRICT, contract N. 320796, and ERC QMAC, contract N. 319286.

\bibliography{biblio}

\end{document}


\centerline{\Huge\textbf{Supplementary Notes}}

\bigskip
\bigskip
\bigskip
In these Supplementary Notes we present in detail the calculations whose results are discussed in the main text, as well as other related ones. 

\section{LUMO and LUMO+1}
The electrons hopping between the sixty carbon atoms of each buckminsterfullerene can be also regarded as 
moving on a sphere and subject to an icosahedral crystal field.  Their wavefunctions are therefore product of a radial one times a combination of spherical harmonics\cite{Trouillier&MartinsPRB1992}. In particular,  
upon defining the real spherical harmonics for $m>0$, 
\be
\begin{split}
Y^c_{lm} &= \fract{1}{\sqrt{2}}\,\Big(Y_{lm} + (-1)^m\,Y_{l-m}\Big)\,,\\
Y^s_{lm} &= -\fract{i}{\sqrt{2}}\,\Big(Y_{lm} - (-1)^m\,Y_{l-m}\Big)\,,
\end{split}\label{real-representation}
\ee
then the angular part of the $t_{1u}$ LUMO is 
\be
\begin{split}
\psi_z\big(t_{1u}\big) &= -\sqrt{\fract{36}{50}}\;Y_{50} - \sqrt{\fract{14}{50}}\;Y^c_{55}\,,\\
\psi_x\big(t_{1u}\big) &=\sqrt{\fract{3}{10}}\;Y^c_{51} -\sqrt{\fract{7}{10}}\;Y_{54}^c\,,\\
\psi_y\big(t_{1u}\big) &=\sqrt{\fract{3}{10}}\;Y^s_{51} +\sqrt{\fract{7}{10}}\;Y_{54}^s\,,
\end{split}\label{t1u}
\ee
while that of the $t_{1g}$ LUMO+1 reads 
\be
\begin{split}
\psi_z\big(t_{1g}\big) &= -Y_{65}^s\,,\\
\psi_x\big(t_{1g}\big) &=\sqrt{\fract{33}{50}}\;Y_{61}^s + \sqrt{\fract{11}{50}}\;Y_{64}^s
+ \sqrt{\fract{6}{50}}\;Y_{66}^s\,,\\
\psi_y\big(t_{1g}\big) &= -\sqrt{\fract{33}{50}}\;Y_{61}^c + \sqrt{\fract{11}{50}}\;Y_{64}^c
- \sqrt{\fract{6}{50}}\;Y_{66}^c\,.
\end{split}\label{t1g}
\ee
These sets of orbitals can be regarded for all purposes as $p$-orbitals. We can also rotate the basis into that of eigenstates of the $z$-component of the $l=1$ angular momentum, 
\be
\begin{split}
\psi_0 &= \psi_z\,,\\
\psi_{+1} &= \fract{1}{\sqrt{2}}\Big(\psi_x+i\psi_y\Big)\,,\\
\psi_{-1} &= -\fract{1}{\sqrt{2}}\Big(\psi_x-i\psi_y\Big)\,.
\end{split}\label{M}
\ee
In what follows we shall use either equivalent representations. Through equations \eqn{t1u} and 
\eqn{t1g} one can easily calculate the angular components of the quadrupole moment 
$Q_u$ and $Q_g$ of the LUMO and LUMO+1, respectively. They have opposite sign and, specifically, 
\be
\fract{Q_g}{Q_u} \simeq -2.6.\label{Q}
\ee
This suggests that the vibronic coupling constants between the $H_g$ modes and the $t_{1g}$ electrons have 
also opposite sign with respect to those of the $t_{1u}$ electrons, and are presumably bigger in absolute value. 

\section{Interaction parameters}

The multipole expansion of the Coulomb interaction 
\[
V(\br,\brp) = \sum_{lm}\,v_l(r,r')\,(-1)^m\,
Y_{lm}\big(\theta_\br,\phi_\br\big)\,Y_{l-m}\big(\theta_\brp,\phi_\brp\big)\,,
\]
projected onto the $t_{1u}$--$t_{1g}$ manifold contains, besides the $l=0$ monopole, i.e. the Slater integral $F_0$ that is simply the Hubbard $U$, an exchange $H_\text{exchange}$, which includes all $l>0$ terms and, implicitly assuming 
normal ordering, can be generally written as 
\be
\begin{split}
H_\text{exchange} =& 
\fract{g_1}{2}\,\Bigg[
\fract{4}{3}\,\sum_a\, n_{u\,a}\,n_{u\,a} - \fract{2}{3}\,\sum_{a\not=b}\,n_{u\,a}\,n_{u\,b} 
+ \sum_{a>b}\,\Delta_{u\,ab}\,\Delta_{u\,ab}\Bigg]\\
& + \fract{g_2}{2}\,\Bigg[
\fract{4}{3}\,\sum_a\, n_{g\,a}\,n_{g\,a} - \fract{2}{3}\,\sum_{a\not=b}\,n_{g\,a}\,n_{g\,b} 
+ \sum_{a>b}\,\Delta_{g\,ab}\,\Delta_{g\,ab}\Bigg]\\
& + g_3\,\Bigg[
\fract{4}{3}\,\sum_a\, n_{u\,a}\,n_{g\,a} - \fract{2}{3}\,\sum_{a\not=b}\,n_{u\,a}\,n_{g\,b} 
+ \sum_{a>b}\,\Delta_{u\,ab}\,\Delta_{g\,ab}\Bigg]\\
& +\fract{g_4}{2}\,n_u\,n_u  +\fract{g_5}{2}\,n_g\,n_g +
g_6\,n_u\,n_g\\
& +\fract{g_7}{2}\,\sum_{ab}\,
\bigg[\Big(\Gamma_{ab} - \Gamma_{ba}\Big)\,\Big(\Gamma_{ab} - \Gamma_{ba}\Big)\bigg]\,,\\
& +\fract{g_8}{2}\,\sum_{ab}\,
\bigg[\Big(\Gamma_{ab} + \Gamma_{ba}\Big)\,\Big(\Gamma_{ab} + \Gamma_{ba}\Big)\bigg]
+\fract{g_9}{2}\,\sum_{ab}\,\Gamma_{aa}\,\Gamma_{bb}\\
&\qquad \equiv \, \sum_{i=1}^9\, H_i\,,
\end{split}\label{exchange}
\ee
where $n_{u\,a}$ and $n_{g\,a}$ are the occupation numbers of orbital $a=x,y,z$ of the $t_{1u}$ LUMO and $t_{1g}$ LUMO+1, respectively, 
$n_{u(g)}=\sum_a\,n_{u(g)\,a}$, while
\ba
\Delta_{u\,ab} &=& \sum_{\sigma}\, \Big(
c^\dagger_{u\,a\sigma}\,c^\dagga_{u\,b\sigma} + H.c.\Big)\,,\\
\Delta_{g\,ab} &=& \sum_{\sigma}\, \Big(
c^\dagger_{g\,a\sigma}\,c^\dagga_{g\,b\sigma} + H.c.\Big)\,,\\
\Gamma_{ab} &=& \sum_{\sigma}\,\Big(
c^\dagger_{u\,a\sigma}\,c^\dagga_{g\,b\sigma} + H.c.\Big)\,,
\ea   
with $c^\dagger_{u(g)\,a\sigma}$ and $c^\dagga_{u(g)\,a\sigma}$
the operators that create and annihilate, respectively, an electron with spin 
$\sigma$ and orbital index $a$ in the $t_{1u(g)}$. 
We mention that $g_1$, $g_2$ and $g_3$ are the coupling constants of the $t_{1u}$-$t_{1u}$, $t_{1g}$-$t_{1g}$ and $t_{1u}$-$t_{1g}$ quadrupole-quadrupole interactions; while $g_7$ the coupling constant of the dipole-dipole interaction. In Table~\ref{interactions} we list the values of the $g_i$'s extracted by models II and III of 
Ref.~\cite{Nikolaev&Michel}. 
\begin{table}
\begin{center}
\begin{tabular}{c||c|c}
 & II & III\\ \hline
$g_1$ & 99.4 & 95.2\\ \hline
$g_2$ & 153.6 &  149.0  \\ \hline 
$g_3$ & -77.2 & -75.0\\ \hline
$g_4$ & 43.7 & 41.7\\ \hline
$g_5$ & 34.3 & 33.7\\ \hline
$g_6$ & 24.7 & 24.7\\ \hline
$g_7$ & 227.9 & 224.7\\ \hline
$g_8$ & 26.3 & 25.0\\ \hline
$g_9$ & -35 & -33.5\\ 
\end{tabular}\qquad 
\begin{tabular}{c||c|c}
 & II & III\\ \hline
$g'_7$ & -724.89 & -710.44\\ \hline
$g'_8$ & 438.17 & 432.83\\ \hline
$g'_9$ & -368.17 & -365.83\\ 
\end{tabular}
\end{center}
\caption{Interaction parameters in meV extracted from 
models II and III of Ref.~\cite{Nikolaev&Michel}.}
\label{interactions}
\end{table}

We note that the Coulomb exchange \eqn{exchange} includes 
a pair-hopping from $t_{1u}$ to $t_{1g}$ and viceversa. 
Since the single particle energy of LUMO+1 lies $\Delta\ep \gtrsim 1.2$~ eV above that of LUMO, we can safely neglect pair-hopping in the calculation of molecular terms. In this approximation and since the $t_{1u}$ and $t_{1g}$ orbitals behave as $p$-orbitals, we can exploit $O(3)$ symmetry. Therefore each state within 
the $\big(t_{1u}\big)^n$ subspace can be labelled by a total angular momentum $L_u$ and its $z$-component $\Lambda_u$, as well as by the total spin $S_u$ and its $z$-component $\Sigma_u$.  Seemingly 
a state within the $\big(t_{1g}\big)^m$ subspace can be labelled by $L_g$, $\Lambda_g$, $S_g$ and 
$\Sigma_g$. It follows that a state in the $\big(t_{1u}\big)^n\otimes \big(t_{1g}\big)^m$ manifold can 
be labelled by total $L$, $\Lambda$, $S$ and $\Sigma$ and defined as 
\be
\begin{split}
\mid L,\Lambda,\,S,\Sigma; \big(n,L_u,S_u\big),\big(m,L_g,S_g\big)\big\rangle =& 
\sum_{\Lambda_u,\Lambda_g,\Sigma_u,\Sigma_g}\,
C^{L\Lambda}_{L_g\Lambda_g\,,L_u\Lambda_u}\, 
C^{S\Sigma}_{S_g\Sigma_g\,,S_u\Sigma_u}\\
& \qquad \qquad \mid m,\,L_g,\Lambda_g,\,S_g,\Sigma_g\big\rangle\,\mid n,\,L_u,\Lambda_u,\,S_u,\Sigma_u\big\rangle\,,
\end{split}
\ee
where $C^{ab}_{cd,ef}$ are Clebsch-Gordan coefficients. The matrix elements of the exchange Hamiltonian, which is a scalar under $O(3)$, can be readily calculated by means of the Wigner-Eckart theorem. \\
In reality, it is convenient to manipulate $H_\text{exchange}$ and make it more manageable through   
the Wigner-Eckart theorem. We define the quadrupole operators 
\[
Q_{2m} = \sqrt{\fract{20\,\pi}{3}\,}\;Y_{2m}\,,
\]
with $m=-2,\dots,2$, for both $t_{1u}$ and $t_{1g}$ electrons. Using the complex representation \eqn{M}, 
so that the orbitals are now labeled by the projection of the $l=1$ angular momentum, i.e. $a=-1,\dots,1$, 
the $m=0$ component of the quadrupole is  
\be
Q_{p\,20} = \fract{1}{\sqrt{3}}\,\Big(2n_{p\,0} - n_{p\,+1} - n_{p\,-1}\Big)\,,
\ee
where $p=u,g$ refers to $t_{1u}$ and $t_{1g}$ electrons. 
One can readily show that, upon transformation into the complex representation \eqn{M}, the following equivalence holds
\be
\begin{split}
&\fract{4}{3}\,\sum_a\, n_{p\,a}\,n_{p\,a} - \fract{2}{3}\,\sum_{a\not=b}\,n_{p\,a}\,n_{p\,b} 
+ \sum_{a>b}\,\Delta_{p\,ab}\,\Delta_{p\,ab} \,\longrightarrow \,
\sum_m\, (-1)^m\, Q_{p\,2m}\,Q_{p\,2-m}\\
& \equiv \mathbb{Q}_{2p}\cdot\mathbb{Q}_{2p}^\dagger 
= 15 - 4S_p\Big(S_p+1\Big) - L_p\,\Big(L_p+1\Big) -\fract{5}{3}\,\Big(n_p-3\Big)^2\,,
\end{split}
\ee
where the dot represents the scalar product between vector spherical harmonics, and 
the last expression is the value on a state identified by $n_p$ electrons in the 
$t_{1p}$ orbital, $p=u,g$, with total angular momentum $L_p$ and spin $S_p$. \\
Moreover the sum of operators in Eq.~\eqn{exchange} that involve the coupling constants $g_7$, $g_8$ 
and $g_9$ can be equivalently written, once pair hopping terms are neglected, as 
\be
\begin{split}
\sum_{i=7}^9\, H_i =& g'_7\,\bigg( \mathbf{S}_u\cdot\mathbf{S}_g +\fract{1}{4}\,n_u\,n_g\bigg)
+ g'_8\,\bigg( \mathbb{Q}_{2g}^\dagger\otimes\mathbf{S}_g
\cdot \mathbb{Q}_{2u}^\dagga\otimes\mathbf{S}_u 
+ \fract{1}{4}\,\mathbb{Q}_{2g}^\dagger
\cdot \mathbb{Q}_{2u}^\dagga\bigg)\\
&  + g'_9\,\bigg(\mathbf{L}_g\otimes\mathbf{S}_g\cdot
\mathbf{L}_u\otimes\mathbf{S}_u +\fract{1}{4}\, 
\mathbf{L}_u\cdot\mathbf{L}_g\bigg)
\, ,
\end{split}\label{exchange-bis}
\ee
where e.g. $\mathbf{L}_u\otimes\mathbf{S}_u$ is the single-particle spin-orbit operator for $t_{1u}$ electrons that corresponds to an orbital operator with $l=1$ and a spin operator with $s=1$. Seemingly 
$\mathbb{Q}_{2u}\otimes\mathbf{S}_u$ is a single-particle operator with orbital momentum $l=2$ 
and spin $s=1$. The new coupling constants are defined in terms of those in Table~\ref{interactions}
through 
\be
\begin{split}
g'_7 &= -\fract{8}{3}\,g_7 -\fract{16}{3}\,g_8 + \fract{2}{3}\,g_9\,,\\
g'_8 &= 2g_7 -2g_8 +g_9\,,\\
g'_9 &= -2g_7 +2g_8 +g_9\,,
\end{split}
\ee
and their values are shown in the same table. 
\\
In conclusion the exchange Hamiltonian \eqn{exchange} without pair hopping terms among $t_{1u}$ and $t_{1g}$ can be rewritten as
\be
\begin{split}
H_\text{exchange} =&\, \fract{g_1}{2}\, \bigg[
15 - 4S_u\Big(S_u+1\Big) - L_u\,\Big(L_u+1\Big) -\fract{5}{3}\,\Big(n_u-3\Big)^2\,\bigg]\\
&+\fract{g_2}{2}\, \bigg[
15 - 4S_g\Big(S_g+1\Big) - L_g\,\Big(L_g+1\Big) -\fract{5}{3}\,\Big(n_g-3\Big)^2\,\bigg]
+ g_3\, \mathbb{Q}_{2g}^\dagger
\cdot \mathbb{Q}_{2u}^\dagga\\
&+\fract{g_4}{2}\,n_u\,n_u  +\fract{g_5}{2}\,n_g\,n_g +
g_6\,n_u\,n_g\\
& + g'_7\,\bigg( \mathbf{S}_u\cdot\mathbf{S}_g +\fract{1}{4}\,n_u\,n_g\bigg)
+ g'_8\,\bigg( \mathbb{Q}_{2g}^\dagger\otimes\mathbf{S}_g
\cdot \mathbb{Q}_{2u}^\dagga\otimes\mathbf{S}_u 
+ \fract{1}{4}\,\mathbb{Q}_{2g}^\dagger
\cdot \mathbb{Q}_{2u}^\dagga\bigg)\\
&  + g'_9\,\bigg(\mathbf{L}_g\otimes\mathbf{S}_g\cdot
\mathbf{L}_u\otimes\mathbf{S}_u +\fract{1}{4}\, 
\mathbf{L}_u\cdot\mathbf{L}_g\bigg)\,,
\end{split}\label{new-exchange}
\ee
which is easier to deal with by means of  Wigner-Eckart theorem, and reproduces all molecular terms obtained in Ref.~\cite{Nikolaev&Michel}. \\
For our purposes, we shall concentrate here only on few configurations. Within the 
$\big(t_{1u}\big)^3$ subspace, we consider all twenty states, i.e.  the multiplets with $L=0$ and $S=3/2$,
$L=2$ and $S=1/2$, and finally $L=1$ and $S=1/2$, which we denote as $^4A_u$, $^2H_u$ and $^2T_{1u}$, respectively. On the contrary, within the 
$\big(t_{1u}\big)^2\big(t_{1g}\big)^1$ subspace, we shall focus only on the states with 
$L=2$ and $S=3/2$, and $L=0$ and $S=3/2$, which we denote as $^4H_g$ and $^4A_g$, respectively.  
Their energies are explicitly 
\be
\begin{split}
E\Big(\,^4A_u\Big) &= -5g_1 + 3g_4\,,\\
E\Big(\,^2H_u\Big) &= -2g_1 + 3g_4\,,\\
E\Big(\,^2T_{1u}\Big) &= 3g_4\,,\\
E\Big(\,^4H_g\Big) &= \Delta\ep -\fract{5}{3}\,g_1 - \fract{1}{3}\,g_3 + g_4 + 2\,g_6 + g'_7 -\fract{1}{6}\,g'_8 +\fract{1}{2}\,
g'_9 \,,\\
E\Big(\,^4A_g\Big) &= \Delta\ep -\fract{5}{3}\,g_1  - \fract{10}{3}\,g_3  +g_4 + 2\,g_6 + g'_7 
-\fract{5}{3}\,g'_8 - g'_9\,,
\end{split}\label{levels}
\ee
where $\Delta\ep$ is the single-particle energy difference between $t_{1g}$ and $t_{1u}$ 
electrons. 
\section{Jahn-Teller effect}
The quadrupole moment of the $t_{1u}$ and the $t_{1g}$ electrons are coupled to the eight fivefold degenerate $H_g$ vibrations, which actually correspond to quadrupolar distortions of the molecule. In 
the single-mode approximation \cite{ManiniPRB98} and using the same conventions, we have to add 
the following operator to the Hamiltonian 
\be
\begin{split}
H_\text{JT} &= \sqrt{\fract{3}{4}\,}\; g\,\omega\,\sum_{m=-1}^1\, (-1)^m\,q_{-m}\,
\Big( Q_{u\,2m} - \lambda\, Q_{g\,2m}\Big) 
+\fract{\omega}{2}\,\Big(\mathbf{q}\cdot\mathbf{q}^\dagger + \mathbf{p}\cdot\mathbf{p}^\dagger\Big)
\,,
\end{split}\label{JT}
\ee
where, as mentioned previously, $\lambda\geq 1$. There are different estimates of $g$ and $\omega$. In 
Table~\ref{gg} we list the results of Ref.~\cite{IwaharaPRB2010} obtained by fitting photoemission spectra 
of C$_{60}^{-}$. 
\begin{table}
\begin{center}
\begin{tabular}{c|c|c|c|c}\hline
~& $g$ & $\omega$(meV)  & $E_\text{JT}=\omega\,g^2/2$ (meV) & $3E_\text{JT}/2$ (meV)\\ \hline
(1) & 1.14 & 100.15 & 64.83 & 97.25\\ \hline
(2) & 1.12 & 99.89 & 62.38 & 93.56\\ \hline
(3) & 1.01 & 96.61 & 57.62 & 86.43\\ \hline
(4) & 1.43 & 76.61 & 78.25 & 117.37\\ \hline
(5) & 1.41 & 82.84 & 82.00 & 123.00\\ \hline
\end{tabular}
\end{center}
\caption{Dimensionless vibronic coupling $g$, phonon frequency $\omega$ and Jahn-Teller energy gain 
$E_\text{JT}=\omega\,g^2/2$ obtained in Ref.~\cite{IwaharaPRB2010} by fitting two sets of photoemission spectra of C$_{60}^{-}$. The labels are the same as in Table I. of \cite{IwaharaPRB2010}. 
(1), (2) and (3) refer to three equally good fits of the same experimental data, while (4) and (5) to fits of another set of data. } \label{gg}
\end{table}

\subsection{Anti-adiabatic regime}
The simplest approximation of the Jahn-Teller Hamiltonian \eqn{JT} is to integrate out the vibrations and neglect the frequency dependence of the vibron-mediated interaction. This corresponds to the so-called anti-adiabatic approximation \cite{nostroRMP}, and amounts to an additional exchange interaction 
\be
\begin{split}
\delta H_\text{exchange} &= -\fract{3}{4}\,E_\text{JT}\, 
\Big(\mathbb{Q}_{u\,2}^\dagger - \lambda\,\mathbb{Q}_{g\,2}^\dagger\Big)\cdot
\Big(\mathbb{Q}_{u\,2}^\dagga - \lambda\,\mathbb{Q}_{g\,2}^\dagga\Big)\,,
\end{split}
\ee
which, unlike Eq.~\eqn{new-exchange}, must not be normal ordered, so that 
\be
\delta H_\text{exchange} = \;:\delta H_\text{exchange}: 
-\fract{5}{2}\,E_\text{JT}\,\Big(n_u+\lambda^2\,n_g\Big)\,,\label{no-normal}
\ee
where $:(\dots):$ denotes normal ordering. 
This implies that the coupling constants $g_1$, $g_2$ and $g_3$ changes according to
\be
\begin{split}
g_1 &\to \, g_{1*} = g_1 - \fract{3}{2}\,E_\text{JT}\,,\\
g_2 &\to \, g_{2*} =g_2 - \fract{3}{2}\,\lambda^2\, E_\text{JT}\,,\\
g_3 &\to \, g_{3*} =g_3 + \fract{3}{2}\,\lambda\, E_\text{JT}\,.
\end{split}\label{change-g}
\ee
Through Eqs.~\eqn{levels} and \eqn{no-normal}, we can actually define three different spin gaps, $\Delta^{(1)}$,  
$\Delta^{(2)}$, and $\Delta^{(3)}$, 
\bea
\Delta^{(1)} &=& E\Big(\,^4A_u\Big)-E\Big(\,^2T_{1u}\Big) = -5\,g_{1*} = 
5\,\Big(\fract{3}{2}\,E_\text{JT}-
g_1\Big)
\simeq -476 + \fract{15}{2}\,E_\text{JT}\,,\label{spin-gap1}\\
\Delta^{(2)} &=& E\Big(\,^4A_g\Big)-E\Big(\,^2T_{1u}\Big) \simeq \Delta\ep-1011 - \fract{5}{2}\,E_\text{JT} \,\Big(2\lambda+\lambda^2-2\Big)\label{spin-gap2}\,,\\
\Delta^{(3)} &=& E\Big(\,^4H_g\Big)-E\Big(\,^2T_{1u}\Big) \simeq \Delta\ep -1135 - \fract{1}{2}\,E_\text{JT} 
\Big(\lambda +5\lambda^2-10\Big)
\,,\label{spin-gap3}
\eea
where energies are in meV and we use the interaction model III of Table~\ref{interactions}. \\

In solution the molecular ground state of C$_{60}^{3-}$ seems to be a spin doublet~\cite{Bhyrappa1993,Reed&Bolskar2000,Boyd1995}, with a very small excitation energy to another magnetic state~\cite{Bhyrappa1993}, which was ascribed to the $^2T_{1u}$ splitting into $^2E_u$ and $^2\!A_u$ in a non-icosahedral environment, or, alternatively, to an excitation between the spin-doublet $^2T_{1u}$ ground state and a spin-quartet with $A$ symmetry~\cite{Green1996}. 
C$_{60}^{2-}$ in solution has instead a non-magnetic ground state with a sizeable spin gap of 75~meV. There are also evidences of a lower magnetic state lying only a wavenumber above the ground state~\cite{Boyd1995}, which might correspond to a genuine C$_{60}^{2-}$ spin excitation or, more likely, to the contribution of C$_{120}$O$^{2-}$ and C$_{120}$O$^{4-}$ impurities~\cite{Reed&Bolskar2000}.\\
In the solid state, NMR spectra in metallic A$_3$C$_{60}$ and 
non-magnetic insulating A$_2$C$_{60}$ reveal the existence of a spin gap of magnitude $75-100\meV$~\cite{AlloulPRB2002,AlloulPRB2002Na2C60}. We must however mention that there are discrepancies between the magnetic susceptibility 
of A$_3$C$_{60}$ (A=K,Rb) measured, e.g., by ESR and by SQUID, see \cite{SQUID} and references therein. While the former is an increasing function of temperature, the latter decreases with increasing $T$. This behaviour has been explained assuming the presence of magnetic impurities~\cite{SQUID}, though it could well indicate the existence of intrinsic low-lying spin excitations.\\ 
Electronic structure 
calculations of isolated molecular anions C$_{60}^{n-}$~\cite{Green1996} find almost vanishing spin gaps between 
distorted $^1\!A_g$ and $^3T_{1g}$ states for $n=2$, and between undistorted $^4\!A_u$ and distorted $^2T_{1u}$ for $n=3$. However, if one adds for $n=3$ the zero-point energy of the molecular vibrations~\cite{Shahab2016}, the energy balance changes appreciably in favour of the distorted spin-doublet, leading to a   
$\Delta^{(1)}=75\meV$, see Eq.~\eqn{spin-gap1}, much in agreement with the experimental value in the solid state. This calculation seems therefore to support the original interpretation of EPR spectra of C$_{60}^{3-}$ in solution given by the authors of Ref.~\cite{Bhyrappa1993}, who associated the 
observed low-lying excitation to the $^2T_{1g}$ splitting in a 
non-icosahedral environment. There are so far no calculations 
of the vibrational zero-point energy contribution in C$_{60}^{2-}$. 
However, since both $^1\!A_g$ and $^3T_{1g}$ allow for unimodal Jahn-Teller distortion, we expect they will have similar zero-point energy gains and thus remain almost degenerate. Therefore, while electronic structure calculations of isolated molecular anions, including the zero-point energy of the molecular vibrations, seem to reproduce 
the physics of C$_{60}^{3-}$ both in solution and in the solid state, 
they might fail in the case of C$_{60}^{2-}$, where instead the evidence of a low-spin ground state well separated from high-spin excited states is more undeniable, especially in alkali fullerides.\\ 
The conventional explanation of this failure invokes the  
screening of Coulomb exchange by nearby polarisable molecules, which 
is absent in single molecule calculations. This argument is in our opinion not fully satisfying. First, we mentioned that single molecule calculations for C$_{60}^{3-}$, including the zero-point energy for vibrations, do predict a spin-doublet ground state with a gap to the spin-quartet state of similar magnitude to that observed in A$_3$C$_{60}$. Second, we do not understand why the 
quadrupole-quadrupole electron-electron interaction should be screened, whereas the interaction between the electron and molecular quadrupoles, of similar origin, should not. We instead tend to believe that electronic structure calculation may rather fail because of  correlation effects not well captured by independent particle schemes.\\
In this perspective, we cannot exclude that the low lying spin excitations observed in C$_{60}^{3-}$ in solutions~\cite{Bhyrappa1993}, and by SQUID measurements in 
K$_3$C$_{60}$ and Rb$_3$C$_{60}$, might actually correspond to a 
genuine molecular excitation between the spin-doublet ground state and a spin-quartet one, different from the $^4\!A_u$ and elusive to electronic structure calculations because of correlations.\\ 
We shall argue that such a state does exist and must be associated to the lowest among $\Delta^{(2)}$ and $\Delta^{(3)}$, i.e. to an excitation between a $\big(t_{1u}\big)^3$ spin-doublet and a $\big(t_{1u}\big)^2\big(t_{1g}\big)^1$ spin-quartet.
Under this assumption, $\Delta^{(1)}\simeq 75\meV$ implies 
\be
E_\text{JT} \gtrsim 74.5\meV\,,\label{value-EJT}
\ee
larger than the estimates (1)--(3) in Table~\ref{gg}, but lower than (4) and (5). 
The magnitudes of $\Delta^{(2)}$ and $\Delta^{(3)}$ critically depend on 
$\Delta\ep$ and $\lambda$. Electronic structure calculations \cite{Green1996} suggests that 
$\lambda\simeq 1.25$. Near IR absorption spectra of C$_{60}^{-}$ \cite{Bolskar1995,TomitaPRL2005} show a main peak at $\Delta\ep_*=1150-1163\meV$, which is the bare $t_{1u}\to t_{1g}$ excitation energy $\Delta\ep$ reduced by a Jahn-Teller contribution. In the antiadiabatic limit we are using here, 
\be
\Delta\ep_* = \Delta\ep -\fract{5}{2}\,E_\text{JT}\,\big(\lambda^2-1\big)\,,
\ee
which, through Eq.~\eqn{value-EJT}, would imply $\Delta\ep=1253-1266\meV$ but negative 
spin gap $\Delta^{(2)}$. This result contradicts the experimental evidence and might be due to the antiadiabatic approximation that overestimates the Jahn-Teller effect, which is stronger in $^4\!A_g$ than in $^2T_{1u}$. 
Since the antiadiabatic approximation has nonetheless the advantage of being very simple and to depend only on the overall Jahn-Teller energy $E_\text{JT}$ and not on the precise values of the vibrational coupling constants to each of the eight $H_g$ modes, we shall keep using such an approximation and cure its deficiency by a 14\% screening reduction 
of the Coulomb exchange that leads to the following estimates
\be
\begin{split}
E_\text{JT} &\simeq 64.6\meV\,,\\
\Delta\ep &\simeq 1241-1254\meV\,,\\
\Delta^{(1)} &\simeq 75\meV\,,\\
\Delta^{(2)} &\simeq 38-51\meV\,,\\
\Delta^{(3)} &\simeq 295-308\meV\,.
\end{split}\label{antiadiabatic-estimate}
\ee

\subsection{Variational calculation}

In order to assess the accuracy of the antiadiabatic limit, in this section we shall attach the Jahn-Teller problem by a variational approach introduced by Wehrli and Sigrist \cite{SigristPRB2007}, which we first briefly sketch.  \\
One starts from the wavefunction of a product state, 
\be
\mid \Psi(\bq)\big \rangle =\mid\bq\rangle _{v}\, 
\otimes \mid\psi(\bq)\rangle _{e}\,,\label{wf-1}
\ee
where $\mid\bq\rangle _{v}$ is a coherent state of the $H_g$ modes with average displacement $\bq$, see Eq.~\eqn{JT}, which serves as a variational parameter, while $\mid\psi(\bq)\rangle _{e}$ is the (Born-Oppenheimer) electronic ground state at fixed $\bq$. 
Since the Hamiltonian has $SO(3)$ symmetry, its expectation value over a wavefunction like \eqn{wf-1} is also invariant under 
rotation of $\bq$. In other words, under the transformation 
\be
\bq =\begin{pmatrix}
q_{+2}\\
q_{+1}\\
q_{0}\\
q_{-1}\\
q_{-2}
\end{pmatrix} 
\to \, \hat{U}(\Theta)\;\bq\qquad 
\mid \Psi(\bq)\big \rangle \to\, \mid \Psi\Big(\hat{U}(\Theta)\;\bq\Big)\,\big \rangle
\,,\label{bq}
\ee   
where $\hat{U}(\Theta)$ is the $SO(3)$ rotation by Euler angle 
$\Theta$, in this specific case 
the Wigner-D matrix with angular momentum $L=2$, the expectation value of the Hamiltonian is independent of $\Theta$.
We can enforce $SO(3)$ symmetry by parametrising 
the displacement $\bq$ through a magnitude $q$ and shape-angle $\alpha$, as 
\[
\bq = 
\fract{q}{\sqrt{2}}\,
\begin{pmatrix}
\sin\alpha\\
0\\
\sqrt{2}\;\cos\alpha\\
0\\
\sin\alpha 
\end{pmatrix}\,,
\] 
and defining a variational wavefunction 
\be
\mid \Psi^L_{MK}(q,\alpha)\,\big\rangle = \mathcal{Q}_{MK}^{L} \mid \Psi\left(\bq\right)\big \rangle \,,\label{wf-2}
\ee
through the projection operator
\[
\mathcal{Q}_{MK}^{L}=\frac{2L+1}{8\pi^{2}}\int d\Theta\, D_{MK}^{L}\left(\Theta\right)\, \hat{U}\left(\Theta\right)\, 
\]
where the integration is over the Euler angles,  
\[
\int d\Theta=\intop_{0}^{2\pi}d\phi\intop_{0}^{2\pi}d\gamma\intop_{0}^{\pi}\sin\theta \,d\theta\,,
\]
$D_{MK}^{L}\left(\Theta\right)$ are the real Wigner-D functions and
$\hat{U}\left(\Theta\right)$ is the rotation operator. 
The variational wavefunction thus depends on two variational parameters, $q$ and $\alpha$, as well as on the quantum numbers $L$, the total angular momentum, its $z$-component $M$, and an additional integer $K$ that is non-zero only in the case of bimodal 
distortions $\sin\alpha\not= 0 \not=\cos\alpha$. \\ 
In the case of $C_{60}^{-}$ with an electron in
the $t_{1u}$ orbital, $\alpha=0$, i.e. the distortion is unimodal, and the variational energy is obtained by minimising
with respect to $q$ the functional 
\[
E\left[\left(t_{1u}\right)^1\right]\left(g,q\right)=\left(\frac{q^{2}}{2}h\left(q\right)-gq\right)\omega
\]
where 
\[
h\left(q\right)\,=\, \fract{\intop_{-1}^{1}t^{2}\left(\frac{3}{2}t^{2}-\frac{1}{2}\right)e^{-\frac{3}{4}q^{2}(1-t^{2})}dt}{\intop_{-1}^{1}t^{2}e^{-\frac{3}{4}q^{2}(1-t^{2})}dt}\;.
\]\\
Seemingly, the energies of $C_{60}^{3-}$ in the $S=1/2$ 
$\big(t_{1u}\big)^3$ subspace, with bimodal distortion and thus $K\not=0$, or in the $S=3/2$ 
$\big(t_{1u}\big)^2\big(t_{1u}\big)^1$ subspace with unimodal distortion, are obtained by minimising with respect to $q$ the functionals
\[
\fract{E\left[\left(t_{1u}\right)^{3}\right]\left(g,q\right)}{\omega}=\mathrm{LowerEigenvalue}\left[\begin{array}{cc}
\fract{q^{2}}{2}\fract{G_{L,K}^{T}}{N_{L,K}^{T}}+\fract{r\,E\left(^{2}T_{1u}\right)}{\omega} & \fract{\sqrt{3}}{2}gq\fract{N_{L,K}^{T}+N_{L,K}^{H}}{\sqrt{N_{L,K}^{T}N_{L,K}^{H}}}\\
\fract{\sqrt{3}}{2}gq\fract{N_{L,K}^{T}+N_{L,K}^{H}}{\sqrt{N_{L,K}^{T}N_{L,K}^{H}}} & \fract{q^{2}}{2}\frac{G_{L,K}^{H}}{N_{L,K}^{H}}+\fract{r\,E\left(^{2}H_{u}\right)}{\omega}
\end{array}\right],
\]
and
\begin{align*}
& \fract{E\left[\left(t_{1u}\right)^{2}\left(t_{1g}\right)^{1}\right]\left(g,q\right)}{\omega}= \Delta\ep\nonumber \\
& \qquad \qquad \qquad  +
\mathrm{LowerEigenvalue}\left[\begin{array}{cc}
\fract{q^{2}}{2}\frac{F_{L}^{1}}{F_{L}^{0}}+\fract{r\,E\left(^{4}A_{g}\right)}{\omega} & \fract{\lambda gq}{\sqrt{2}}\left(\sqrt{\frac{F_{L}^{1}}{F_{L}^{0}}}+\sqrt{\fract{F_{L}^{0}}{F_{L}^{1}}}\right)\\
\fract{\lambda gq}{\sqrt{2}}\left(\sqrt{\fract{F_{L}^{1}}{F_{L}^{0}}}+\sqrt{\fract{F_{L}^{0}}{F_{L}^{1}}}\right) & \fract{q^{2}}{2}\fract{F_{L}^{2}}{F_{L}^{1}}-\lambda gq+\fract{r\,E\left(^{4}H_{g}\right)}{\omega}
\end{array}\right],
\end{align*}
where
\begin{eqnarray}
N_{L,K}^{T}\left(q\right) & = & \frac{2L+1}{8\pi^{2}}\int d\Theta\, D_{KK}^{L}(\Theta)\,
D_{00}^{1}(\Theta)\;
\text{e}^{-\frac{q^{2}}{2}\big(1-D_{22}^{2}(\Theta)\big)}\;,\nonumber \\
N_{L,K0}^{H}\left(q\right) & = & \frac{2L+1}{8\pi^{2}}\int d\Theta\,
 D_{KK}^{L}(\Theta)\, D_{-2-2}^{2}(\Theta)\, 
 \text{e}^{-\frac{q^{2}}{2}\big(1-D_{22}^{2}(\Theta)\big)}\;,\nonumber \\
G_{L,K}^{T}\left(q\right) & = & \frac{2L+1}{8\pi^{2}}\int d\Theta\, D_{KK}^{L}(\Theta)\, 
D_{00}^{1}(\Theta)\, D_{22}^{2}(\Theta)\,\text{e}^{-\frac{q^{2}}{2}\big(1-D_{22}^{2}(\Theta)\big)}
\;,\nonumber \\
G_{L,K}^{H}\left(q\right) & = & \frac{2L+1}{8\pi^{2}}\int d\Theta\, D_{KK}^{L}(\Theta)\, 
D_{-2-2}^{2}(\Theta)\, D_{22}^{2}(\Theta)\, \text{e}^{-\frac{q^{2}}{2}\big(1-D_{22}^{2}(\Theta)\big)}
\;,\nonumber \\
F_{L}^{n}\left(q\right)&=&\frac{2L+1}{2}\intop_{-1}^{1}dt\, P_{L}\left(t\right)\, \Big(P_{2}\left(t\right)\Big)^{n}\,\text{e}^{-\frac{3}{2}q^{2}\big(1-t^{2}\big)}\;,
\end{eqnarray}
involve real Wigner-D functions and Legendre Polynomials. 
The
lower energies are obtained for $\left(L,K\right)= \left(1,0\right),\left(2,-2\right)$, and for $L=0,2$
in the $\left(t_{1u}\right)^{3}$ 
and $\left(t_{1u}\right)^{2}\left(t_{1g}\right)^{1}$ cases, respectively. \\
We have optimised the wavefunctions for two vibrational frequencies in the single-mode approximation, rows (1) and (4) in Table~\ref{gg}. The value of $\Delta\ep$ is obtained as before through the main near-IR absorption peak $\Delta\ep_* = 1150-1163\meV$ subtracting the Jahn-Teller contribution. In the variational scheme the latter can be approximately obtained through the product state \eqn{wf-1} assuming a vertical Franck-Condon transition, which provides a somehow lower estimate $\Delta\ep\simeq 1080\meV$. 
We observe that the two different vibrational frequencies lead to similar results when plotted as function of the Jahn-Teller energy. Moreover, consistently with the Jahn-Teller effect being overestimated within the anti-adiabatic approximation, in this variational approach we get sensible results only assuming a larger screening reduction, $r=0.78$ of the bottom panels, and larger Jahn-Teller energy $E_\text{JT}\sim 90\meV$. 
Such a value is above the estimates extracted from most recent photoemission data, (1)--(3) in Table~\ref{gg}, but close to those usually adopted in the literature, see e.g. \cite{SigristPRB2007}. 
Moreover, as suggested in Refs.~\cite{BunnPRB2008,DunnPRB2013}, 
quadratic couplings among the electrons and $H_g$ modes, allowed since $H_g\times H_g$ includes still $H_g$ and not accounted for in the fit, might be not negligible. 
In view of the uncertainty in the values of the vibrational coupling constants and in the accuracy of the single-mode approximation, in the main text we have preferred to emphasise more the results obtained in the antiadiabatic approximation, which depend on a single vibrational parameter $E_\text{JT}$. 
\begin{figure}[H]
\centerline{\includegraphics[width=13cm]{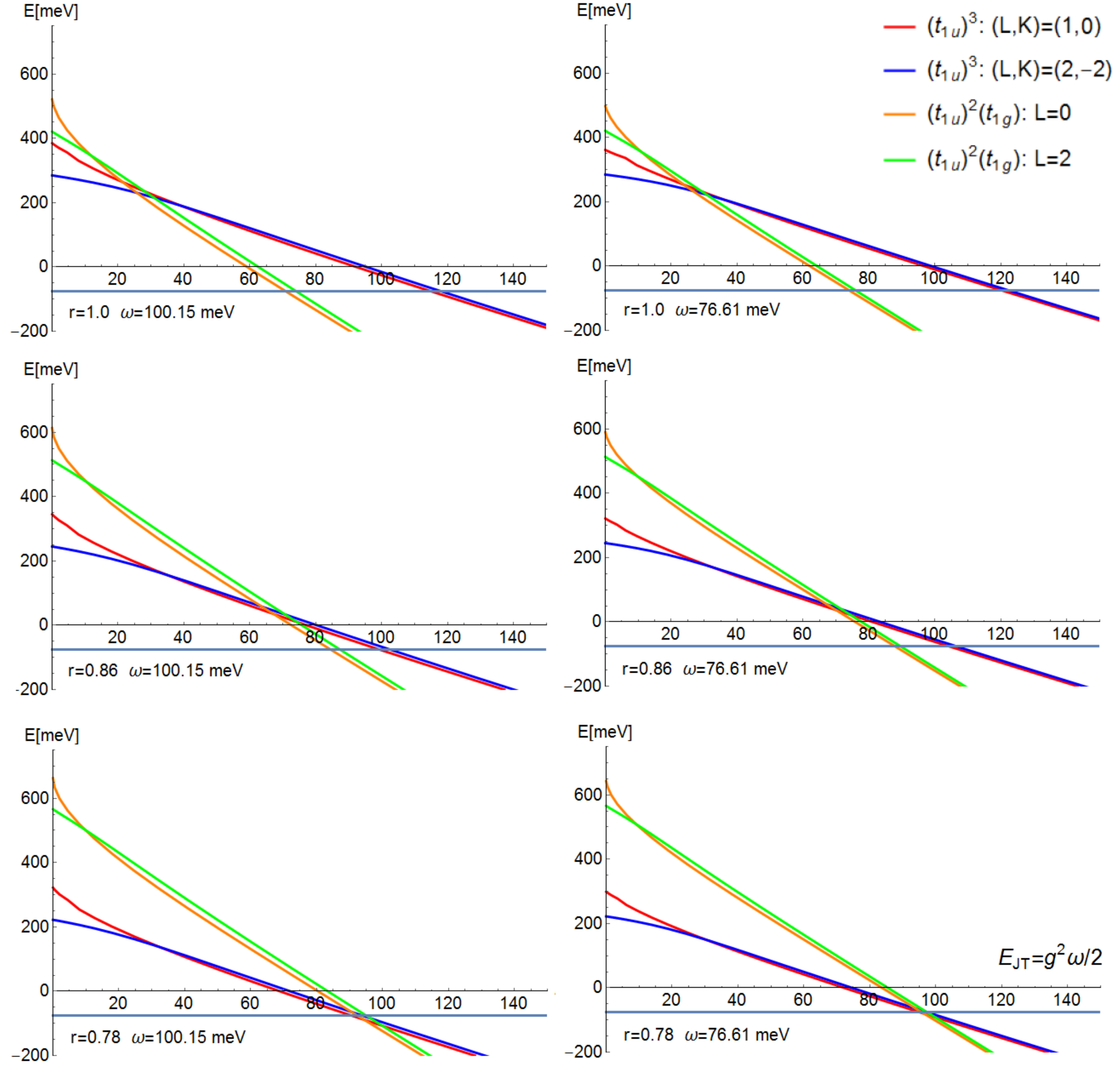}}
\caption{Variational energies for the $\left(t_{1u}\right)^{3}$  and the $\left(t_{1u}\right)^{2}\left(t_{1g}\right)^{1}$ configurations. Left panels are obtained with $\omega=100.13\meV$ meV 
while right panels with $\omega=76.61\meV$. Top panels corresponds to $r=1$, central panels to $r=0.86$, 
and finally bottom panels to $r=0.78$. The zero of energy is fixed at the energy 
of the $^4\!A_u$ state, while the horizontal line at $E=-75\meV$ is supposedly the energy of the 
$^2T_{1u}$, which corresponds to $(L,K)=(1,0)$. Seemingly, the state $^2H_u$ corresponds to 
$(L,K)=(2,-2)$, while $L=0$ and $L=2$ correspond to the states $^4\!A_g$ and $^4H_g$, respectively.}
\label{energie}
\end{figure}

\begin{figure}[H]
\vspace{-2cm}
\centerline{\includegraphics[width=12cm]{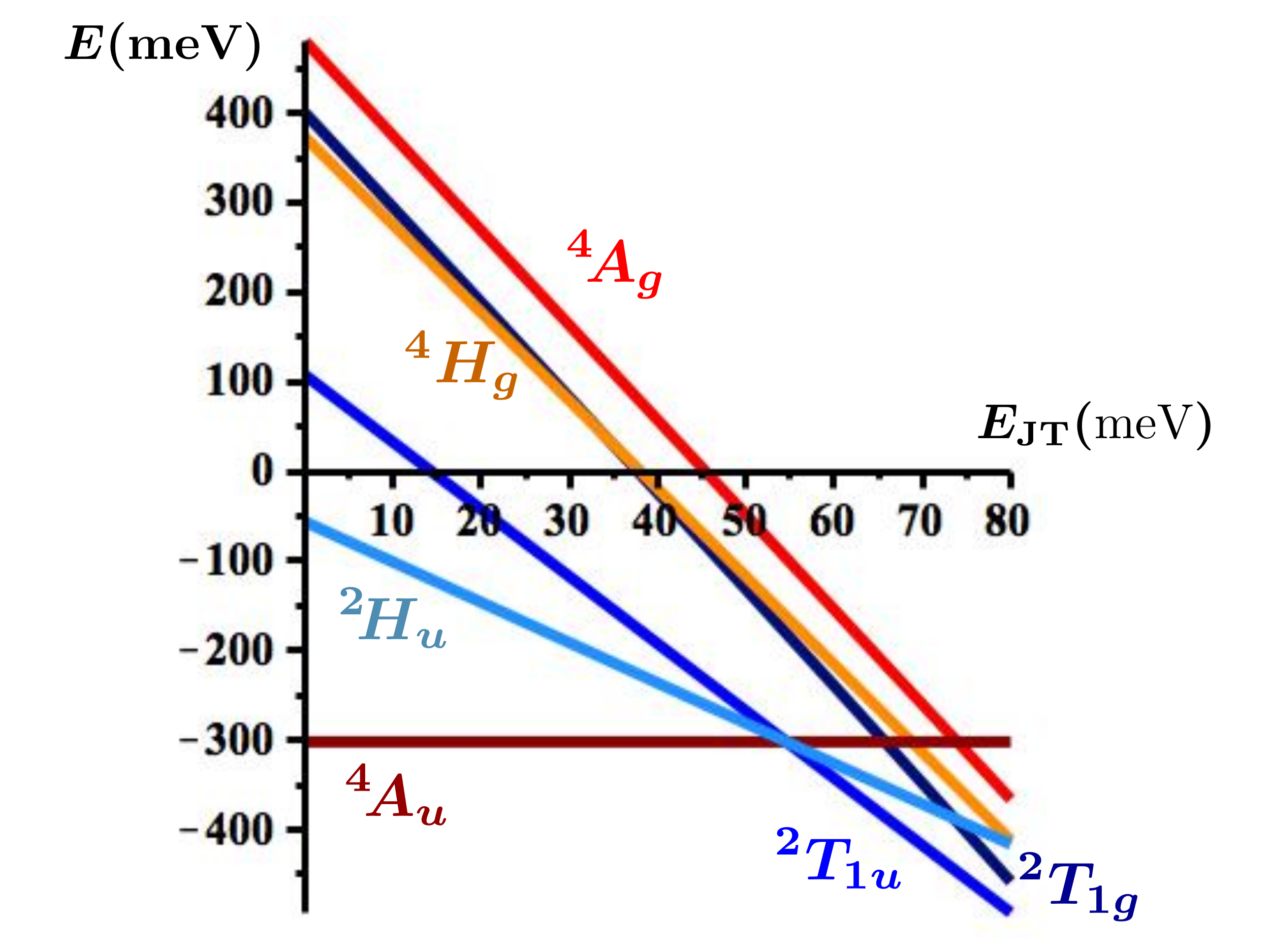}}
\caption{Molecular terms calculated in the antiadiabatic approximation, to be compared with Fig. 2(a) of the manuscript. The parameters are the same as in that figure, but replacing Eq.~\eqn{change-g} with \eqn{change-g-2} with $\lambda_2=0.1$ and 
$\lambda_1=1.7$. The dark blue line is a $^2T_{1g}$ state pushed by the new parameters that we use close to the $^2T_{1u}$ ground state.}
\label{livelli}
\end{figure}
We conclude with some comments concerning the results of the present and the previous sections. Originally we were hoping that, in the presence of Jahn-Teller,  the lowest energy state within the 
$\big(t_{1u}\big)^2\big(t_{1g}\big)^1$ subspace should have been one within the  manifold of symmetry $^2T_{1g}$, which includes a very low-lying state already in the absence of JT. Evidently that circumstance would have been very pleasant since the $^2T_{1u}\to ^2T_{1g}$ transition is parity and spin allowed. \\
We mention that the  $^2T_{1g}$ manifold comprises three states: a state (1) obtained by the $\big(t_{1u}\big)^2$ $^1\!A_g$ plus one electron in the $t_{1g}$; a state (2) obtained by the $\big(t_{1u}\big)^2$ $^1\!H_g$ coupled to one electron in the $t_{1g}$ into a 
$^2T_{1g}$-symmetry state; and finally a state (3) obtained by the $\big(t_{1u}\big)^2$ $^3\!T_{1g}$ coupled to one electron in the $t_{1g}$ still into a $^2T_{1g}$-symmetry state. We note 
that the $^1\!A_g$ state within the $\big(t_{1u}\big)^2$ subspace is the one gaining the largest Jahn-Teller energy. Therefore our expectation was that $^1\!A_g \oplus \big(t_{1g}\big)^2 \to ^2T_{1g}(1)$ should have 
been the winner among the three. However, the Coulomb exchange splitting couples among each other all the three $^2T_{1g}$ states, more strongly the state (2) with (3). As the result, the lowest $^2T_{1g}$ state in the absence of JT is a mixture with almost similar weights of (2) and (3) and none of (1). The next, 1~eV above, is mostly contributed by (1), and lastly the highest, 2~eV above, is again a mixture mainly of (2) and (3). \\
The Jahn-Teller effect, though sizeable, is not strong enough to push (1) down enough and below the spin-quartet $^4\!A_g$. However there might still be a chance for that to happen under particular circumstances that would need further investigation. It has been suggested~\cite{BunnPRB2008} that the $t_{1g}$ electrons may couple preferentially of $H_g$ modes that are instead the less coupled to $t_{1u}$ electrons. If this were confirmed, one should better work in a two-mode approximation, rather than a single-mode. Alternatively, within the anti-adiabatic approximation,  the equation \eqn{change-g} should be replaced by 
\be
\begin{split}
g_1 &\to \, g_{1*} = g_1 - \fract{3}{2}\,E_\text{JT}\,,\\
g_2 &\to \, g_{2*} =g_2 - \fract{3}{2}\,\lambda_1^2\, E_\text{JT}\,,\\
g_3 &\to \, g_{3*} =g_3 + \fract{3}{2}\,\lambda_2\, E_\text{JT}\,.
\end{split}\label{change-g-2}
\ee
with $0<\lambda_2\ll \lambda_1$. With properly massaged values of $\lambda_2=0.1$ and 
$\lambda_1=1.7$ we obtain the molecular terms of Fig.~\ref{livelli}, to be compared with Fig.~2(a) of the manuscript. Indeed with such values we are able to push $^2T_{1g}$ below $^4\!A_g$ and close to 
the $^2T_{1u}$ ground state. As we mentioned, this scenario could explain much more simply the origin of the mid infrared peak. However the cooling mechanism that we propose, see the following section, would not work anymore. One could consider a different scheme taking into account the radiative decay of such an exciton, which would be much faster than that of $^4\!A_g$, and the fact that also the $^2T_{1g}$ is 
entropy-rich. We tend however to believe more in the physical parameters used in the manuscript, mainly because this new scenario requires $\lambda_2\lesssim 0.1$, which implies that the set of $H_g$ modes that couple to $t_{1u}$ has almost no overlap to that coupling to $t_{1g}$. Nonetheless we think this alternative possibility worth to be further investigated by more \textit{ab-initio} techniques.

\section{Boltzmann equation and effective temperature}

In this section we give some details on the Boltzmann equation that we use to 
describe the time evolution of the quasiparticle distribution under
the effect of the laser pulse. First we make explicit all approximations. We mentioned that the energy supplied by the laser pulse is partly used to create the exciton by absorbing/emitting spin-triplet quasiparticle-quasihole excitations, and partly to emit $H_g$ vibrations because of the Franck-Condon effect. 
We shall neglect the latter and thus assume a $\delta$-like spectrum of the exciton 
$\mathcal{A}_\text{exc}(\ep) = \delta\big(\ep-E_\text{exc}\big)$. Within such an approximation, the effect of the laser is therefore: (1) to create excitons and correspondingly lower the density of quasiparticles; (2) change the distribution of spin-triplet quasiparticle-quasihole excitations. The diminishing of quasiparticle density below half-filling already brings about a reduction of quasiparticle entropy. We shall neglect this effect and just concentrate on the distribution of quasiparticle-quasihole excitations, which implies underestimating the entropy reduction. 
Finally, we shall not take into account the radiative decay of an exciton back into a photon with simultaneous emission/absorption of a spin-triplet quasiparticle-quasihole excitation, as well as non-radiative decay processes. \\
The equations that control the rate of creation of triplet excitons per molecule, $N_{S_z}$ with $S_z=-1,0,1$,  are 
\be
\begin{split}
\dot{N}_{+1}(t) =& \,\gamma\,\Big(1-N_\text{exc}(t)\Big)\,\sum_{a=1}^3\, \int d\ep \,\rho(\ep)\,\rho(\ep+\Delta\omega)\,
n_{a\up}(\ep,t)\,\bar{n}_{a\down}\big(\ep+\Delta\omega,t\big)\Big)\,,\\
\dot{N}_{-1}(t) =& \,\gamma\,\Big(1-N_\text{exc}(t)\Big)\,\sum_{a=1}^3\, \int d\ep \,\rho(\ep)\,\rho(\ep+\Delta\omega)\,
n_{a\down}(\ep,t)\,\bar{n}_{a\up}\big(\ep+\Delta\omega,t\big)\Big)\,,\\
\dot{N}_{0}(t) =& \, \fract{\gamma}{2}\,\Big(1-N_\text{exc}(t)\Big)\,\sum_{a=1}^3\,\sum_\sigma\,\, \int d\ep \,\rho(\ep)\,\rho(\ep+\Delta\omega)\,
n_{a\sigma}(\ep,t)\,\bar{n}_{a\sigma}\big(\ep+\Delta\omega,t\big)\Big)\,,
\end{split}\label{B-1}
\ee
where $n_{a\sigma}(\ep,t)$ is the distribution of quasiparticles 
with orbital index $a$ and spin $\sigma$, 
$\bar{n}_{a\sigma}(\ep,t)=1-n_{a\sigma}(\ep,t)$, $\Delta\omega=\omega-E_\text{exc}$, with $\omega$ the laser frequency, 
and $\rho(\ep)$ the quasiparticle DOS.  The factor $\Big(1-N_\text{exc}(t)\Big)$, where $N_\text{exc}(t)=N_{+1}(t)+N_0(t)+N_{-1}(t)$, 
stems from the fact that only one exciton per molecule 
can be created. The coupling constant $\gamma$ depends on the strength of the absorption process as well as on the properties of the light beam, and can be parametrised as 
\be
\gamma = \alpha\, \mathcal{N}(\omega)\,\hbar \omega\,,\label{gamma-alpha}
\ee
where $\mathcal{N}(\omega)$ is the density of photons per molecule and $\alpha$ a parameter of dimension $\big[\text{time}^{-1}\big]$ that depends only on the system properties. Correspondingly, the equation of motion of the quasiparticle distribution is  
\be
\begin{split}
\dot{n}_{a\up}(\ep,t) =& \,\fract{\gamma}{2}\,\Big(1-N_\text{exc}(t)\Big)\,\bigg[
-2\,n_{a\up}(\ep,t)\,\bar{n}_{a\down}\big(\ep+\Delta\omega,t\big)
-n_{a\up}(\ep,t)\,\bar{n}_{a\up}\big(\ep+\Delta\omega,t\big)\\
&\qquad +2\,\bar{n}_{a\up}(\ep,t)\,n_{a\down}\big(\ep-\Delta\omega,t\big)
+\bar{n}_{a\up}(\ep,t)\,n_{a\up}\big(\ep-\Delta\omega,t\big)\bigg]+I_{a\up}(\ep,t)\,,\\
\dot{n}_{a\down}(\ep,t) =&\, \fract{\gamma}{2}\,\Big(1-N_\text{exc}(t)\Big)\,\bigg[
-2\,
n_{a\down}(\ep,t)\,\bar{n}_{a\up}\big(\ep+\Delta\omega,t\big) -n_{a\down}(\ep,t)\,\bar{n}_{a\down}\big(\ep+\Delta\omega,t\big)\\
&\qquad +2\,\bar{n}_{a\down}(\ep,t)\,n_{a\up}\big(\ep-\Delta\omega,t\big)\ +\bar{n}_{a\down}(\ep,t)\,n_{a\down}\big(\ep-\Delta\omega,t\big)\bigg]+I_{a\down}(\ep,t)\,,
\end{split}\label{B-2}
\ee
where $I_{a\sigma}(\ep,t)$ is the collision integral due to the residual interaction among quasiparticles. By writing Eq.~\eqn{B-1}
and Eq.~\eqn{B-2} we have assumed that the spin-triplet quasiparticle-quasihole spectrum is that of weakly interacting quasiparticles. In other words, the paramagnon enhancement is taken into account only in the strength of the parameter $\alpha$ and not in the spectral redistribution of the joint quasiparticle-quasihole density of states. \\
Since the collision integrals conserve the quasiparticle energy 
per molecule, $\mathcal{E}(t)$, one can readily verify that the rate of its change is simply
\be
\dot{\mathcal{E}}(t) = \sum_{a\sigma}\,\int d\ep\,\ep\,\rho(\ep)\,
\dot{n}_{a\sigma}(\ep,t) = 
\Big(\omega-E_\text{exc}\Big)\,\dot{N}(t)\,.\label{B-3}
\ee
On the contrary, the number of quasiparticles per molecule 
$\mathcal{N}(t)$ is assumed to be constant, even though, in reality, 
$\dot{\mathcal{N}}(t)=-\dot{N}_\text{exc}(t)$. Eq.~\eqn{B-3} implies that 
below resonance, i.e. for $\omega<E_\text{exc}$, the energy of quasiparticles diminishes, while above it increases. We emphasise that this result is due to our approximation $\mathcal{A}_\text{exc}(\ep) = \delta\big(\ep-E_\text{exc}\big)$ and to neglecting the reduction in quasiparticle number. \\
We solve the Boltzmann equations \eqn{B-1} and \eqn{B-2} assuming 
that the collision integral is so strong that local equilibrium is established at any instant of time during the pulse duration. We can therefore define an instantaneous value of the temperature $T(t)$. If we assume, besides the conservation of quasiparticle number, also particle-hole symmetry, then the chemical potential remains zero during the evolution. Under these assumptions it follows that 
\be
n_\sigma(\ep,t) = f\big(\ep,T(t)\big) \equiv \bigg(1+\exp\fract{\ep}{T(t)}\bigg)^{-1}\;,\label{n-f}
\ee
namely the distribution at time $t$ is the Fermi-Dirac one at temperature $T(t)$. Therefore, through equations 
\eqn{B-1} and \eqn{B-3}, we have to solve
\be
\begin{split}
\dot{\mathcal{E}}(t) =& \, 6\,\fract{\dot{T}(t)}{T(t)^2}\int d\ep\,\ep^2\,\rho(\ep)\, 
 f\Big(\ep,T(t)\Big)\bigg[1-f\Big(\ep,T(t)\Big)\bigg]\\
 =& \,9\gamma\,\Delta\omega\,\int d\ep\,\rho(\ep)\,\rho(\epsilon+\Delta\omega)\, 
 f\Big(\ep,T(t)\Big)\,
 \bigg[1-f\Big(\ep+\Delta\omega,T(t)\Big)\bigg]\,,
 \end{split}\label{dot-T}
 \ee
which is a first order integro-differential equation for the instantaneous temperature $T(t)$ with 
initial condition $T(0)=T$, where $T$ is the equilibrium value of the temperature before the pulse. This equation can be integrated numerically. In particular we used a Runge-Kutta based algorithm, and  
a semicircular $\rho(\ep)$ with half-bandwidth of $50\meV$.  The 
effective temperature at the end of the pulse with duration $300$~fs is therefore 
$T_\text{eff} = T\big(t=300~\text{fs}\big)$.
 \\
In order to make contact with optical conductivity, we calculate in linear response the energy absorbed 
by the quasiparticle plus the exciton at frequency $\omega$ per unit time, in seconds, and per molecule, which, through \eqn{B-1}
and \eqn{B-3} and relaxing the assumption 
$\mathcal{A}_\text{exc}(\ep) = \delta\big(\ep-E_\text{exc}\big)$, 
reads 
\be
\begin{split}
\mathcal{W}(\omega) =& \,\hbar\omega\,\dot{N}(t)
\simeq 9\,\gamma\,\hbar\omega\,\int d\ep_\text{exc}\,\mathcal{A}_\text{exc}\big(\ep_\text{exc}\big)\, \int d\ep\,
\rho(\ep)\,\rho(\ep+\omega-\ep_\text{exc})\,
f(\ep,T)\,\bar{f}\big(\ep+\omega-\ep_\text{exc},T\big)\\
&\equiv 9\,\alpha\,\hbar^2\,\omega^2\,\mathcal{N}(\omega)\,
J(\omega,T)
\,,
\end{split}
\ee
where $\bar{f}(\ep,T)=1-f(\ep,T)$.  The density of the electromagnetic field at frequency $\omega$ and per molecule is 
\[
E_\text{EMF}(\omega) = \hbar\omega\,
\mathcal{N}(\omega)\,,
\]
so that, since in the units of Ref.~\cite{Cavalleri2016}
\[
1\,\Omega\,\text{cm} = 0.423\,\text{ps}\;\rightarrow 
\fract{\hbar}{\Omega\,\text{cm}} = 1.56\meV
\,,
\]
 the absorption rate in unit energy is 
\be
\fract{\hbar}{\tau(\omega)} \simeq 1.56\meV\,\sigma_1(\omega)\big[\Omega^{-1}\,\text{cm}^{-1}\big]= 
\fract{\hbar\mathcal{W}(\omega)}{E_\text{EMF}(\omega)} = 
9\,\hbar\alpha\,\hbar\omega\,J(\omega,T)
\,.
\ee
\begin{figure}[H]
\centerline{\includegraphics[width=7cm]{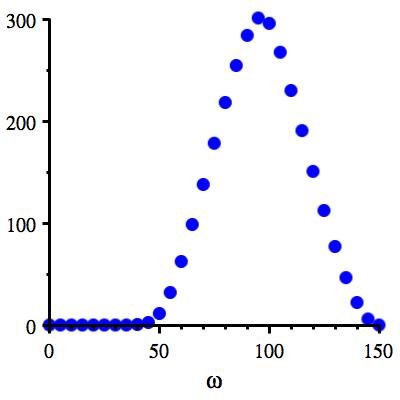}
\hspace{1cm}\includegraphics[width=7cm]{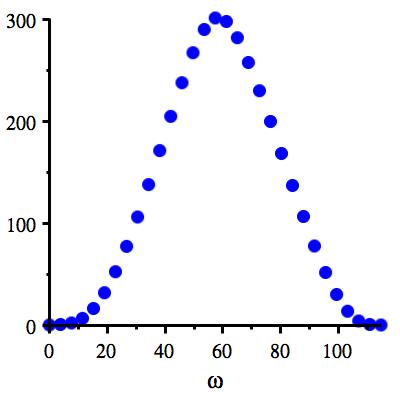}
}
\caption{Calculated contribution of the excitonic absorption to optical conductivity at $T=25$~K 
assuming semicircular $\rho(\ep)$ and $\mathcal{A}_\text{exc}(\ep)$, the former with half-bandwidth 
$50\meV$, left panel, and $40\meV$, right panel, and the latter with a tenth of it. The exciton is centred at 50~meV, left panel, and 
20~meV, right panel.}
\label{sigma1}
\end{figure}
In the left panel of Fig.~\ref{sigma1} we plot the excitonic contribution to optical conductivity 
calculated at $T=25$~K assuming semicircular $\rho(\ep)$ and $\mathcal{A}_\text{exc}(\ep)$, the former with half-bandwidth 
$D=50\meV$ and the latter with a tenth of it and centred at $E_\text{exc}=50\meV$. The fitting parameter is such as to give a peak of $300\big[\Omega^{-1}\,\text{cm}^{-1}\big]$, which we find corresponds to $\hbar\alpha=2.2\,D$. In comparison with the experimental data, the curve is shifted to higher frequency. A better fit could be obtained with smaller values of 
$E_\text{exc}=20\meV$ and $D=40\meV$, right panel of the figure. We repute those values a bit unphysical and believe that the higher peak frequency as compared with experiment is rather due to our approximation of the dynamical spin susceptibility with that of weakly interacting quasiparticles. \\
The experiment of Ref.~\cite{Cavalleri2016} is performed at fixed fluence, which implies at fixed value of $E_\text{EMF}(\omega)=\hbar\omega\,\mathcal{N}(\omega)$. 
A fluence of $1~\text{mJ}\,\text{cm}^{-2}$ with a penetration depth of 
$220\text{~nm}$~\cite{Cavalleri2016} would correspond to an energy of the electromagnetic field per molecule of around $E_\text{EMF}(\omega)=207\meV$. Given the crudeness of our modelling and the neglect of other absorption processes, in the calculation we took 
a smaller $E_\text{EMF}(\omega)= 100\meV$, which provides results closer to the experiment.



\bibliographystyle{unsrt}
\bibliography{biblio}